\documentclass[usenatbib,useAMS]{mn2e}
\bibpunct{(}{)}{;}{a}{}{,}
\usepackage{mn2e-breakabs}
\usepackage[dvips]{graphicx}
\usepackage{amsmath}
\usepackage{amssymb} 
\usepackage{subfigure}
\usepackage{url}
\def\rpet{\mbox{$r_{\rm petro}$}}

\def\zphot{\mbox{$z_{\rm phot}$}}
\def\zspec{\mbox{$z_{\rm spec}$}}
\begin{document}

\title[GAMA: Galaxy clustering using photometric redshifts]
{Galaxy and Mass Assembly (GAMA): Colour and luminosity dependent clustering 
from calibrated photometric redshifts}

\author[L.~Christodoulou et al.]{
\parbox{\textwidth}{\raggedright
L.~Christodoulou$^{1}$\thanks{Email: L.Christodoulou@sussex.ac.uk},
C.~Eminian$^{1}$,
J.~Loveday$^{1}$,
P.~Norberg$^{2}$, 
I.K.~Baldry$^{3}$,  
P.D.~Hurley$^{1}$,
S.P.~Driver$^{4,5}$,
S.P.~Bamford$^6$,
A.M.~Hopkins$^7$,
J.~Liske$^{8}$,
J.A.~Peacock$^9$,
J.~Bland-Hawthorn$^{10}$, 
S.~Brough$^7$,
E.~Cameron$^{11}$,
C.J.~Conselice,$^6$
S.M.~Croom$^{10}$,
C.S.~Frenk$^{2}$,
M.~Gunawardhana$^{10}$, 
D.H.~Jones$^{12}$,
L.S.~Kelvin$^{4,5}$,
K.~Kuijken$^{13}$,
R.C.~Nichol$^{14}$,
H.~Parkinson$^{9}$,
K.A.~Pimbblet$^{12}$,
C.C.~Popescu$^{15}$,
M.~Prescott$^{3}$,
A.S.G.~Robotham$^{4,5}$,
R.G.~Sharp$^{16}$,
W.J.~Sutherland$^{17}$,
E.N.~Taylor$^{18}$,
D.~Thomas$^{14}$,
R.J.~Tuffs$^{19}$,
E.~van~Kampen$^8$,
D.~Wijesinghe$^{10}$
}
\vspace{0.4cm}\\
\parbox{\textwidth}{
$^{1}$ Astronomy Centre, University of Sussex, Falmer, Brighton BN1 9QH, UK\\
$^{2}$ Institute for Computational Cosmology, Department of Physics,
Durham University, South Road, Durham DH1 3LE, UK\\
$^{3}$ Astrophysics Research Institute, Liverpool John Moores University,
$^{4}$ ICRAR (International Centre for Radio Astronomy Research), University of Western Australia, Crawley, WA6009, Australia\\
$^{5}$ SUPA (Scottish Universities Physics Alliance), School of Physics \& Astronomy, University of St Andrews, North Haugh, St Andrews, Fife, KY169SS, UK \\
Twelve Quays House, Egerton Wharf, Birkenhead, CH41 1LD, UK\\
${^6}$ Centre for Astronomy and Particle Theory, University of Nottingham, University Park, Nottingham NG7 2RD, UK\\
$^{7}$ Australian Astronomical Observatory, P.O. Box 296, Epping, NSW 1710, Australia\\
$^{8}$ European Southern Observatory, Karl-Schwarzschild-Str.~2, 85748
Garching, Germany\\
$^{9}$ Institute for Astronomy, University of Edinburgh, 
Royal Observatory, Blackford Hill, Edinburgh EH9 3HJ, Scotland\\
$^{10}$ Sydney Institute for Astronomy, School of Physics, University of
Sydney, NSW 2006, Australia\\
$^{11}$ Department of Physics, Swiss Federal Institute of Technology (ETH-Z{\" u}rich), 8093 Z{\" u}rich, Switzerland\\
$^{12}$ School of Physics, Monash University, Clayton, Victoria 3800,
Australia\\
$^{13}$ Leiden University, P.O.~Box 9500, 2300 RA Leiden, The Netherlands\\
$^{14}$ Institute of Cosmology and Gravitation (ICG), University of
Portsmouth, Dennis Sciama Building, Burnaby Road, Portsmouth PO1 3FX,
UK\\
$^{15}$ Jeremiah Horrocks Institute, University of Central Lancashire,
Preston PR1 2HE, UK\\
$^{16}$ Research School of Astronomy \& Astrophysics, Mount Stromlo Observatory, Weston Creek, ACT 2611, Australia\\
$^{17}$ Astronomy Unit, Queen Mary University London, Mile End Rd, London
E1 4NS, UK\\
$^{18}$ School of Physics, University of Melbourne, Victoria 3010, Australia\\
$^{19}$Max Planck Institute for Nuclear Physics (MPIK), Saupfercheckweg
1, 69117 Heidelberg, Germany
}}
%*****************************************************************************

\maketitle

\begin{abstract}
We measure the two-point angular correlation function of a sample of 4,289,223 galaxies with $r < 19.4$ mag from the Sloan Digital Sky Survey as a function of photometric redshift, absolute magnitude and colour down to
$M_r - 5 \log h = -14$ mag. Photometric redshifts are estimated from $ugriz$ model magnitudes and two Petrosian radii using the artificial neural network package ANNz, taking advantage of the Galaxy and Mass Assembly (GAMA) spectroscopic sample as our training set. The photometric redshifts are then used to determine absolute magnitudes and colours. For all our samples, we estimate the underlying redshift and absolute magnitude distributions using Monte-Carlo resampling. These redshift distributions are used in Limber's equation to obtain spatial correlation function parameters from power law fits to the angular correlation function. 
We confirm an increase in clustering strength for sub-$L^*$ red galaxies compared with $\sim L^*$ red galaxies  at small scales in all redshift bins, whereas for the blue population the correlation length is almost independent of luminosity for $\sim L^*$ galaxies and fainter. A linear relation between relative bias and log luminosity is found to hold down to luminosities $L\sim0.03L^*$. We find that the redshift dependence of the bias of the $L^*$ population can be described by the passive evolution model of \citet{TegmarkPeebles1998}. A visual inspection of a random sample of our $r < 19.4$ sample of SDSS galaxies reveals that about 10 per cent are spurious, with a higher contamination rate towards very faint absolute magnitudes due to over-deblended nearby galaxies.  We correct for this contamination in our clustering analysis. 

\end{abstract}

%*****************************************************************************


%*****************************************************************************

\section{Introduction}

Measurement of galaxy clustering is an important cosmological tool in
understanding the formation and evolution of galaxies at different
epochs.  
The dependence of galaxy
clustering on properties such as morphology, colour, luminosity or
spectral type has been established over many decades. Elliptical
galaxies or galaxies with red colours, which both trace an old stellar
population, are known to be more clustered than spiral galaxies
(e.g.~\citealt{Davis1976,Dressler1980,Postman1984,Loveday1995,Guzzo1997,Goto2003}). Recent large galaxy surveys have allowed the investigation of
galaxy clustering as a function of both colour and luminosity (\citealt{Norberg2002,Budavari2003,zehavi05,Wang2007,McCracken2008,Zehavi2010}). Among the red
population, a strong luminosity dependence has been observed whereby 
luminous galaxies are more clustered, because they reside in
denser environments.   

The galaxy luminosity function shows an increasing faint-end density to
at least as faint as $M_r - 5 \log h = -12$ mag (\citealt{Blanton2005, Loveday2012}), thus
intrinsically faint galaxies represent the majority of the galaxies in the universe. 
These galaxies with luminosity $L \ll L^*$ have low stellar mass and are mostly dwarf
galaxies with ongoing star formation. However, because most wide-field
spectroscopic surveys can only probe luminous galaxies over
large volumes, this population is often under-represented.
Previous clustering analyses have revealed that intrinsically
faint galaxies have different properties to luminous ones. 
A striking difference appears between galaxy colours in this regime:
while faint blue galaxies seem to cluster on a scale almost
independent of luminosity, the faint red population is shown to be
very sensitive to luminosity
\citep{Norberg01,Norberg2002,Zehavi02,Hogg2003,zehavi05,Swanson2008,Zehavi2010,Ross2011}. As found by
\citet{zehavi05}, this trend is naturally explained by the halo
occupation distribution framework. In this picture, the faint red
population corresponds to red satellite galaxies, which are
located in high mass halos with red central galaxies and are
therefore strongly clustered. Recently, \citet{Ross2011} compiled from the literature bias measurements for red
galaxies over a wide range of luminosities for both spectroscopic and
photometric data. They showed that the bias measurements of the faint
red population are strongly affected by non-linear effects and thus on
the physical scales over which they are measured. They conclude that red galaxies with $M_r>-19$ mag are similarly or less biased than red galaxies of intermediate luminosity.

In this work, we make use of photometric redshifts to probe the
regime of intrinsically faint galaxies. Our sample is composed of
SDSS galaxies with $r$-band Petrosian magnitude $r_{\text{petro}}<19.4$. As we have an ideal
training set for this sample, thanks to the GAMA survey \citep{Driver2010},
we use the artificial neural network package ANNz \citep{ANNz}
to predict photometric redshifts. We then calculate
the angular two-point correlation function as a function of absolute
magnitude and colour. The correlation length of each sample is
computed through the inversion of Limber's equation, using Monte-Carlo resampling for modelling the underlying redshift distribution. Recently, \cite{Zehavi2010} presented the clustering properties of the DR7 spectroscopic sample of SDSS. They extracted a sample of $\sim$ 700,000 galaxies with redshifts to $r\leq17.6$ mag, covering an area of 8000 $\text{deg}^2$. Their study of the luminosity and colour dependence uses power law fits to the projected correlation function. Our study is complementary to theirs, since we are using calibrated photo-$z$s of fainter galaxies from the same SDSS imaging catalogue. We use similar luminosity bins to Zehavi et al., with the addition of a fainter luminosity bin $-17 < M_r - 5\log h < -14$.

Small-scale $(r<0.1h^{-1}\text{Mpc})$ galaxy clustering provides additional tests of the fundamental problem of how galaxies trace dark matter. Previous studies have used SDSS data and the projected correlation function to study the clustering of galaxies at the smallest scales possible \citep{Masjendi2006}, using extensive modeling to account for the fibre constraint in SDSS spectroscopic data. The interpretation of these results offers unique tests about how galaxies trace dark matter and the inner structure of dark matter halos \citep{Watson2011}. Motivated by these  studies we present measurements of the angular correlation function down to scales of $\theta \approx 0.005$ degrees. We work solely with the angular correlation function and we pay particular attention to systematics errors and the quality of the data.  

On the other hand, on sufficiently large scales ($r > 60 \ h^{-1} \text{Mpc}$), it is expected that the galaxy density field evolves linearly following the evolution of the dark matter density field \citep{Tegmark06}. However, it is less clear if this assumption holds on smaller scales, where complicated physics of galaxy formation and evolution dominate. In the absence of sufficient spectroscopic data to comprehensively study the evolution of clustering, \cite{Ross2010} used SDSS photometric redshifts to extract a volume-limited sample with $M_r<-21.2$ and $\zphot<0.4$. Their analysis revealed significant deviations from the passive evolution model of \citet{TegmarkPeebles1998}. Here we perform a similar analysis, again using photometric redshifts, for the $L^*$ population.  

This paper is organised as follows. In Section~\ref{clust:2pcorr}, we introduce the statistical quantities to calculate the clustering of galaxies, with an emphasis on the angular correlation function. In Section~\ref{clust:data} we present our data for this study and the method for estimating the clustering errors. In Section~\ref{clust:photoz} we describe the procedure that we followed in order to obtain the photometric redshifts. We then investigate the clustering of our photometric sample, containing a large number of intrinsically faint galaxies, in Section~\ref{clust:corres}. In Section~\ref{clust:bias} we present bias measurements as functions of colour, luminosity and redshift. Our findings are summarised in Section~\ref{clust:concl}. In Appendix~\ref{clust:query} we show how we extracted our initial catalogue from the SDSS DR7 database and finally in Appendix~\ref{clust:systematics} we describe in some detail the tests performed to assess systematic errors.  

Throughout we assume a standard flat $\Lambda$CDM cosmology, with $\Omega_m=0.30$, $\Omega_{\Lambda}=0.70$ and $H_0 = 100h$ km s$^{-1}$ Mpc$^{-1}$. 

%*****************************************************************************

\section{The two-point angular correlation function}\label{clust:2pcorr}

\subsection{Definition}\label{clust:def}

The simplest way to measure galaxy clustering on the sky is via
the two-point correlation function, $w(\theta)$, which gives the excess
probability of finding two galaxies at an angular separation
$\theta$ compared to a random Poisson distribution \citep[\S~31]{peebles80}:

\begin{equation}
dP = \bar{n}^{2}[1+w(\theta)] d\Omega_{1} d\Omega_{2},
\end{equation}
where $dP$ is the joint probability of finding galaxies in solid
angles $d \Omega_{1}$ and $d \Omega_{2}$ separated by $\theta$,
and $\bar{n}$ is the mean number of objects per solid angle. If
$w(\theta)$ $=$ 0, then the galaxies are unclustered and
randomly distributed at this separation. We consider various
estimators for $w(\theta)$ in Section~\ref{clust:estimator}.

\subsection{Power law approximation}\label{clust:powlaw}

Over small angular separations, the two-point correlation function can be approximated by a power law:

\begin{equation}\label{equ:powlaw_w}
w(\theta) = A_{w} \theta^{1-\gamma},
\end{equation}
where $A_{w}$ is the amplitude. The amplitude of the correlation function of a galaxy population is reduced as we go to higher redshifts, because equal angular separations trace larger spatial separations for more distant objects. 
By contrast, the slope $1-\gamma$, of the correlation function is observed to
vary little from sample to sample, with $\gamma \approx 1.8$.
It is mostly sensitive to galaxy colours (see
Section~\ref{clust:corres}).

\subsection{Estimator}\label{clust:estimator}

In practice, the calculation of $w(\theta)$ is
done through the normalised counts of galaxy-galaxy pairs
$DD(\theta)$ from the data, random-random pairs $RR(\theta)$ from an unclustered random catalogue which follows the survey angular selection function, and galaxy-random
pairs $DR(\theta)$.
Various expressions have been used to calculate $w(\theta)$.
In this work we adopt the estimator introduced by
\cite{Landy1993}, which is widely used in the literature:

\begin{equation}\label{equ:wls}
w(\theta) = \frac{DD(\theta) - 2DR(\theta) +
RR(\theta)}{RR(\theta)}.
\end{equation}
\cite{Landy1993} showed that this estimator has a small variance, close to Poisson, and allows one
to measure correlation functions with minimal uncertainty and bias.
The counts $DD(\theta)$, $DR(\theta)$ and $RR(\theta)$ have to be normalised
to allow for different total numbers of galaxies $n_{g}$ and
random points $n_{r}$:

\begin{eqnarray*}
DD(\theta) &=& \frac{N_{gg}(\theta)}{n_{g}(n_{g}-1)/2},\nonumber\\
DR(\theta) &=& \frac{N_{gr}(\theta)}{n_{g}n_{r}},\\
RR(\theta) &=& \frac{N_{rr}(\theta)}{n_{r}(n_{r}-1)/2}.\nonumber\\
\end{eqnarray*}

We use approximately ten times as many random points as galaxies in order that the results
do not depend on a particular realization of random distribution. We also tried an alternative estimator proposed by \cite{hamilton93} which revealed no significant changes in the correlation function measurements.  

Estimates of the angular correlation function are affected by an integral constraint of the form
\begin{equation}\label{equ:ic}
\frac{1}{\Omega^2}\iint w(\theta_{12}) d \Omega_1 d \Omega_2 = 0,
\end{equation}
where the integral is over all pairs of elements of solid angle $\Omega$, within the survey area.
The constraint requires that $w(\theta)$ goes negative at large separations, 
to balance the positive clustering signal at smaller separations.
However, for wide-field surveys like SDSS the integral constraint has a negligible effect on $w(\theta)$,
even on large scales.
We find that the additive correction for the integral constraint is at least two order of magnitude smaller than the value of $w(\theta)$ at $\theta = 9.4$ degrees. Thus the integral constraint does not bias our clustering measurements. 

\subsection{Spatial correlation function}\label{clust:xi}

We are interested in the spatial clustering and the
physical separations at which galaxies are clustered, in order to compare data against theory. To this end, we need
to calculate the spatial correlation function from our angular
correlation function, which is simply its projection on the sky.
The spatial correlation function, $\xi(r)$, can be also expressed
as a power law
\begin{equation}\label{equ:powlaw_xi}
\xi(r) = \left(\frac{r}{r_{0}}\right)^{-\gamma},
\end{equation}
where $r_{0}$ is the correlation length. It
corresponds to the proper separation at which the probability of
finding two galaxies is twice that of a random
distribution, $\xi(r_{0})=1$. 
\citet{Limber1953} demonstrated that the power law approximation
for $\xi(r)$ in equation~\ref{equ:powlaw_xi} leads to the power law
defined in equation~\ref{equ:powlaw_w} with the index $\gamma$ being the same in both cases. \citet{Phillipps1978} expressed the amplitude of the correlation function, $A_{w}$, as a function of the proper correlation length, $r_{0}$, and of the selection function of the survey, whereas later studies propose similar equations where the selection function is implicitly included in the redshift distribution.

Now, writing the angular correlation function as
$w(\theta)=A_{w}\theta^{1-\gamma}$, Limber's
equation becomes  \citep[\S~52, 56]{peebles80}:
\begin{equation}
A_w = C\frac{\int_{z_{min}}^{z_{max}} r_{0}^{\gamma}g(z) (dN/dz)^2 dz}{[\int_{z_{min}}^{z_{max}} (dN/dz) dz]^2},
\label{equ:limber_inv}
\end{equation}
where $dN/dz$ is the redshift distribution\footnote{We use the expressions $dN/dz$ and $N(z)$ interchangeably for the redshift distribution.}, which is zero everywhere outside the limits $z_{min}$ and $z_{max}$ and
\begin{equation}\nonumber
C=\pi^{1/2} \frac{\Gamma[(\gamma-1)/2]}{\Gamma(\gamma/2)}
\end{equation}
with $\Gamma$ the gamma function. The quantity $g(z)$ is
defined as
\begin{equation}\nonumber
g(z) = \left(\frac{dz}{dx}\right) x^{1-\gamma}F(x)
\end{equation}
where $F(x)$ is related to the curvature factor $k$ in the
Robertson-Walker metric by:
\begin{equation}\nonumber
F(x) = 1 - kx^2.
\end{equation}
We assume zero curvature, and so $F(x) \equiv 1$.

When using equation~\ref{equ:limber_inv}, we need to determine the redshift
distribution of the sample with precision. We address this issue in Section~\ref{clust:dndz}. Another subtle complication which arises from the use of equation~\ref{equ:limber_inv} is that galaxy clustering is assumed to be independent of galaxy properties such as colour and luminosity \citep[\S~51]{peebles80}. Therefore it is particularly important to use samples with fixed colour and luminosity, instead of mixed populations for studying galaxy clustering using Limber's approximation. We address this issue in Section~\ref{clust:bins} where we define the colour and luminosity bins for the clustering analysis. 

\section{Data}\label{clust:data}

To carry out this analysis, we take advantage of the Galaxy
and Mass Assembly (GAMA) survey \citep{Driver2010}. This
spectroscopic sample, at low to intermediate redshifts, forms an ideal
training set for predicting photometric redshifts of faint
galaxies. The galaxies considered for the calculation of the
correlation functions are drawn from the seventh data release of the
Sloan Digital Sky Survey photometric sample (SDSS DR7;
\citealt{DR7}). We briefly outline the properties
of these samples below.

\subsection{SDSS DR7 photometric sample}
At the time of writing, the Sloan Digital Sky Survey
(SDSS) is the largest local galaxy survey ever undertaken.
The completed SDSS maps almost one quarter of the sky, with optical
photometry in $u$, $g$, $r$, $i$ and $z$ bands and spectra
for $\sim10^6$ galaxies. The main goal of the survey is to
provide data for large-scale structure studies of the local universe. A series of papers describe the survey: technical information about the data products and the pipeline can
be found in \citet{York2000} and in \citet{Stoughton2002}. Details
about the photometric system can be found in \citet{Fukugita1996}.

The SDSS imaging survey is completed with the seventh data
release \citep{DR7}, that we use in this paper. The main program of SDSS is
concentrated in the Northern Galactic cap with three $2.5^{\circ}$ stripes in the Southern Galactic cap. SDSS DR7 contains about $5.5\times10^6$ galaxies with $r_{\rm petro} < 19.4$
over 7,646 $\text{deg}^2$ of sky.

The images are obtained with a 2.5-meter telescope,
located at Apache Point Observatory, New Mexico.
Various flux measures are available for
galaxies in the SDSS database \citep{Stoughton2002}, including Petrosian
fluxes, model fluxes (corresponding to whichever of a de Vaucouleurs or 
exponential profile provides a better fit to the observed galaxy profile), 
and aperture fluxes. 
In this paper we use model magnitudes to calculate galaxy
colours and Petrosian magnitudes to split galaxies in absolute
magnitude ranges. After \citet{Schlegel1998}, we correct the magnitudes with dust attenuation
corrections provided for each object and each filter in the SDSS database.

The star-galaxy classification adopted by the SDSS photometric
pipeline is based on the difference between an object's PSF magnitude
(calculated assuming a point spread function profile, as for
a stellar source) and its model magnitude. An
object is then classified as a galaxy if it satisfies the criterion
\citep{Stoughton2002}
\begin{equation}\label{class_star_gal}
m_{\rm psf,tot} - m_{\rm model,tot} > 0.145,
\end{equation}
where $m_{\rm psf,tot}$ and $m_{\rm model,tot}$ magnitudes are obtained
from the sum of the fluxes over $ugriz$ photometric bands. This
cut works at the 95 per cent confidence level for galaxies with $r<21$. In Section~\ref{sec:gama} we discuss a different star-galaxy classification, following the GAMA survey, which is the one we adopt for this work (see also Appendix~\ref{clust:query}). 

A photometric redshift study can be vulnerable to contamination not only due to stars misclassified as galaxies, but also to contamination due to over-deblended sources \citep{Scranton02}, usually coming from local spiral galaxies. This imposes limits on the angular scale over we can probe the correlation function. In order to test for this systematic in our sample, in Appendix~\ref{clust:systematics_faint} we visually inspect random samples of the data and then we model the contamination as a function of angular separation. 

\subsection{GAMA sample}\label{sec:gama}

The Galaxy and Mass Assembly (GAMA) 
project\footnote{\url{http://www.gama-survey.org}} 
is a combination of several ground and space-based surveys
with the aim of improving our understanding of galaxy formation
and evolution \citep{Driver2010}. 
GAMA uses the AAOmega spectrograph of the Anglo-Australian
Telescope (AAT) for spectroscopy \citep{Saunders2004,Sharp2006}. Its targets are selected from the SDSS photometric sample. Target selection is described in detail by \cite{Baldry2010}. The main restriction is that the source
is detected as an extended object: $r_{\rm psf}-r_{\rm model}>0.25$. As shown in Appendix~\ref{clust:query}, this criterion is also adopted for our sample extraction from SDSS. This criterion is more restrictive, in the sense that fewer stars will be 
mis-classified as galaxies, than the star-galaxy
classification adopted by the SDSS photometric pipeline (previous
Section), but similar to that used for the SDSS main galaxy
spectroscopic sample \citep{Strauss2002}.

The GAMA survey is almost 99 per cent spectroscopically complete over its 144 deg$^2$ area to
$\rpet = 19.4$ mag \citep{Driver2010}. GAMA phase 1 (comprising 3 years of observations)  %(tilingcat v09) 
includes 95,592 reliable spectroscopic galaxy redshifts to this magnitude limit, extending to redshift $z \approx 0.5$. Of these redshifts, 76,360 have been newly-acquired by the GAMA team. The rest come from previous surveys: SDSS \citep{DR7}, 2dFGRS \citep{Colless2001,Cole2005}, 6dFGS \citep{Jones2004}, MGC \citep{Driver2005} and 2SLAQ \citep{cannon06}. The overall GAMA redshift distribution is shown in Fig.~13 of \citet{Driver2010}. 

For a consistent training of ANNz it is necessary to match all the GAMA objects with SDSS DR7 \"ubercal photometry \citep{Padmanabhan2008} and perform identical colour cuts. Once we apply the colour cuts (Section~\ref{clust:photoz_colourcuts}) necessary for the optimization of ANNz performance, and low and high redshifts cuts ($0.002<z<0.5$), 93,584 redshifts remain. They are used to train our photometric redshift neural net algorithm as described in Section~\ref{clust:photoz}.

\subsection{Colour cuts}\label{clust:photoz_colourcuts}
Before we build our final sample from ANNz, we remove galaxies
with outlier $u-g$, $g-r$, $r-i$, $i-z$ colours both in the SDSS imaging sample and in the training set, because photometric redshift estimates are
based primarily on these colours. The complete colour and magnitude cuts are given in Table~\ref{table:colour_cuts}. Less than $1$ per cent of the galaxies are affected by the colour cuts. These colour cuts in principle could affect the mask that we use for correlation function calculations. To estimate the extent of this effect we study the distribution on the sky of the colour outliers as well as their angular correlation function. This exercise reveals that colour outliers have a spurious correlation an order of magnitude larger on all angular scales than the correlation function of our final sample. 
However, since the number of these objects is almost three orders of magnitude less than the total, they would have a negligible effect on $w(\theta)$ measurements if included.

\begin{table}
\caption{Colour and apparent magnitude cuts for the optimization of ANNz. All magnitudes are SDSS model magnitudes.}
\label{table:colour_cuts}
%\vspace{2mm}
\begin{center}
\begin{math} 
\begin{array}{c}
\hline\hline
12.0<\rpet< 19.4 \\
%0<m_{\text{tot}}< 30 \\
-2<u-g< 7 \\
-2<g-r< 5 \\
-2<r-i< 5 \\
-2<i-z< 5 \\
\hline
\end{array}
\end{math}
\end{center}
\end{table}

\subsection{Final sample}\label{clust:sample}
Our aim is to obtain a galaxy sample with photometric properties
as close as possible to our training set. To this end, we have
selected galaxies from the SDSS DR7 photometric sample with the
query used to select GAMA targets (Appendix~\ref{clust:query}). We select galaxies which have ``clean'' photometry according to the instructions given on the SDSS website\footnote{\url{http://www.sdss.org/dr7/products/catalogs/flags.html}}. Our
sample is hence limited by $r_{\rm petro}<19.4$ and satisfies the
criterion for star-galaxy separation $r_{\rm psf}-r_{\rm
model}>0.25$. In our analysis, we choose to calculate the
correlation function for galaxies located in the SDSS northern
cap, corresponding to 92 per cent of SDSS DR7 galaxies. As such, the
geometry of the survey is simplified to a contiguous area.
Our final sample, after the colour cuts given in Table~\ref{table:colour_cuts} comprises 4,890,965 galaxies. 

To evaluate the number of data-random and random-random pairs in
equation~\ref{equ:wls}, we need to build a mask for our sample.
The mask precisely defines the sky coverage of the sample. We
use the file \verb|lss_combmask.dr72.ply| in the NYU Value
Added Catalogue\footnote{\url{http://sdss.physics.nyu.edu/vagc/}}
\citep{Blanton2005b},
mapping SDSS stripes, as our mask. This file contains the
coordinates of the fields observed by SDSS expressed in spherical
polygons, excluding areas around bright stars because galaxies in these regions can be affected
by photometric errors. It is also suitably formatted for use with the {\sc mangle} software \citep{hamilton93,Hamilton2004,Swanson2008b}, a tool for manipulating survey masks and obtaining random points with the exact geometry of the mask. Once masking is applied, 4,511,011 galaxies remain in our sample. 

The upper panel of Fig.~\ref{fig:ra_dec_jk} shows the boundaries of the final mask for SDSS DR7 that we use for creating random catalogues. Our random catalogues consist of $\sim 10^7$ objects, approximately ten times larger than the number of galaxies in each luminosity and colour bin. Consistency checks have shown that our clustering results are not sensitive to any particular realization of the random catalogue.  In Appendix~\ref{clust:scaling} we check the accuracy of the survey mask, as well as the photometric uniformity of the sample, by studying the angular clustering of our sample as a function of $r$-band apparent magnitude. 

\subsection{Pixelisation scheme and jackknife resampling}\label{clust:pix}
\begin{figure}
\begin{center}
\includegraphics[width=1.\linewidth]{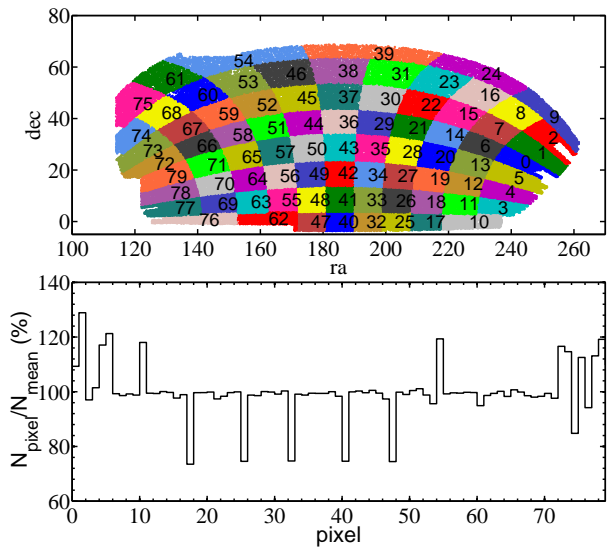}
\caption{The upper panel shows the jackknife regions used for the error estimation of our correlation function measurements. After modifying the SDSSPix scheme, there are 80 jackknife regions which contain approximately equal numbers of random points. The lower panel reports the normalized area of each pixel, based on a random catalogue. The deviations from uniformity show that differences in the areas of the JK regions are limited to $\pm30$ per cent at most. } 
\label{fig:ra_dec_jk}
\end{center}
\end{figure}

In order to speed up the computation of the correlation function,
we pixelise our data according to the
SDSSPix\footnote{\url{http://dls.physics.ucdavis.edu/~scranton/SDSSPix/}}
scheme. The basic concept consists of assigning galaxies located in a portion of the sky to a pixel.
After this step, we only need to take into account galaxies in the
same pixel and in the neighbouring pixels to calculate the
correlation function up to the scale of a pixel. SDSSPix divides
the sky along SDSS $\eta$ and $\lambda$ spherical coordinates 
(as defined in Section 3.2.2 of
\citealt{Stoughton2002}) in equal spherical areas. Different
resolutions are available according to the angular scale of
interest. We choose the resolution called basic resolution %\emph{superpixel}
(resolution $=1$). This divides the sky in 468 pixels of size $\sim 9.4\times 9.4$
deg. Then, for galaxies in a given pixel, that pixel and its 8
direct neighbouring pixels include all neighbouring galaxies with
separations up to 9.4 degrees, the largest angular separation we consider
(see Section~\ref{clust:corres}). 

We also use this pixelisation scheme to define the Jackknife (JK) regions
for the error analysis. In order to minimize the variation in the
number of objects in each JK region, some neighbouring pixels that
contain the survey boundary are
merged in order that they contain a more nearly equal number of random points.
This modification of the SDSSPix pixelisation yields 80 JK regions, as shown in the upper panel of Fig.~\ref{fig:ra_dec_jk}. The lower panel of Fig.~\ref{fig:ra_dec_jk} presents the relative variation in area of each region, as measured by the relative number of randoms each one contains. Hereafter, errors on $w(\theta)$ are determined from 80 JK resamplings, by calculating $w(\theta)$ omitting each region in turn. We have checked that our results are not significantly affected by using either 104 or 40 Jackknife regions. The elements of the covariance matrix, $\boldsymbol{C}$, are given by:
\begin{equation}\label{eq:cov}
C_{ij} = \frac{N-1}{N}\sum^N_{k=1}(\text{log}(w^k_i)-\text{log}(\bar{w_i}))(\text{log}(w^k_j)-\text{log}(\bar{w_j})),
\end{equation}
where $w^k_i$ is the angular correlation function of the $k^{\rm th}$ JK resampling on scale $\theta_i$, $\bar{w_i}$ the mean angular correlation function and N the total number of JK resamplings. In practice, $\bar{w_i}$ is identical with the angular correlation function measurement from the whole survey area. The $N-1$ factor in the numerator of equation~\ref{eq:cov} accounts for correlations inherent in the jackknife procedure \citep{Miller1974}.

Jackknife is a method of calculating uncertainties on a quantity that that we measure from the data itself. In wide-field galaxy surveys, more often than not, large superstructures appear to significantly influence clustering measurements. The best known example is the SDSS Great Wall \citep{Gott2005}. The presence of such structures makes it tempting to present the results with and without the JK region that encloses them, as done in the clustering studies of \cite{zehavi05,Zehavi2010}. Better still, \cite{Norberg2011} devise a more objective method to consistently remove outlier JK regions, from the distribution of all JK measurements that one has at hand. We follow that method in the present analysis, and find that for all samples considered, the number of JK regions that are outliers, and therefore removed, is mostly two or three and no more than five.  

\section{Photometric redshifts}\label{clust:photoz}

\begin{figure}
\begin{center}
\includegraphics[width=\linewidth]{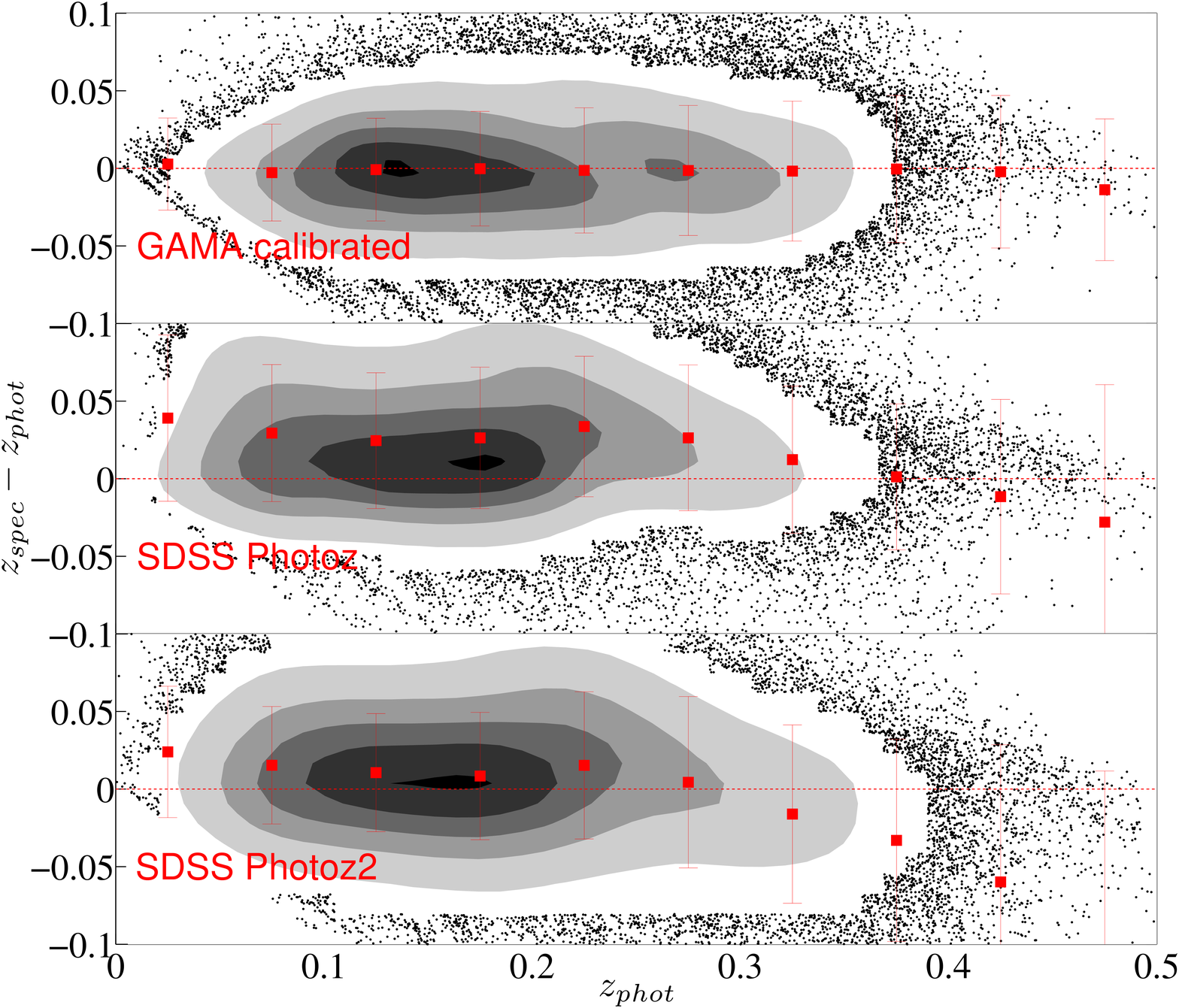}
\caption{Density/scatter plot of redshift error (spectroscopic minus photometric redshift) against predicted photo-$z$ from this work (top panel) and SDSS (middle and bottom panels). The colour coding is such that the densest area (black contour) is 5 times denser than the white contour.  Points are drawn whenever the density of points is less than 10 per-cent of the maximum (black contour). The red squares and error bars represent the mean redshift errors and their standard deviations in photo-$z$ bins of width $\Delta z_{\text{phot}}=0.05$. Horizontal red lines show the zero error benchmark. The improvement in photometric redshift estimates in this work, due primarily to use of the representative GAMA training set, is clear.
}
\label{fig:zspec_vs_zphot}
\end{center}
\end{figure}

For the clustering measurements presented in this paper, all distance information comes from photometric redshifts (photo-$z$). Photo-$z$s are the basis for estimating the redshift distributions to be used in equation~\ref{equ:limber_inv} and in estimating distance moduli to calculate absolute magnitudes and colours. For this study we have a truly representative subset of SDSS galaxies down to $r<19.4$ and we therefore use the artificial neural network package ANNz developed by \citet{ANNz} to obtain photo-$z$ estimates. 

It is important that the training set and the final galaxy sample from SDSS are built using the same selection criteria. The input parameters are the following: \"ubercalibrated, extinction-corrected model magnitudes in $ugriz$ bands,
the radii enclosing 50 per cent and 90 per cent of the Petrosian $r$-band flux of the galaxy, and their respective
uncertainties. The architecture of the network is 7:11:11:1, with seven input parameters described above, two hidden layers with 11 nodes each and a single output, the photo-$z$. We use a committee of 5 networks to predict the photo-$z$s and their uncertainties (see Section~\ref{clust:photoz_errors}).

\subsection{Photometric redshift errors}\label{clust:photoz_errors}

Before we proceed with the photo-$z$ derived quantities that we use in
this study, we investigate the possible biases and errors that ANNz
introduces, using the known redshifts from GAMA. Following standard
practice we split our data into three distinct sets: the training set,
the validation set and the test set. Half of the objects constitute
the test set and the other two quarters the training and validation
sets. This investigation is insensitive to the exact numbers in these
three sets. The training and validation sets are used for training the network, whereas the test set is treated as unknown. Given predicted photo-$z$s $z_\text{{phot}}$, we can quantify the redshift error 
for each galaxy in the test set as
\begin{equation}\label{eq:errors}
\delta z \equiv z_{\text{spec}} - z_\text{{phot}},
\end{equation}
the primary quantity of interest as far as true redshift errors are concerned. It can depend on apparent magnitude, colour, the output $z_\text{{phot}}$, the intrinsic scatter $z_\text{{err}}$ of ANNz committees, as well as the position of an object on the sky if the survey suffers from any photometric non-uniformity. We investigate some of these potential sources of error below. The dispersion $\sigma_z$, of $\delta z$ is given by the equation 
\begin{equation}
\sigma_z^2=\left<\left(\delta z\right)^2\right>- \left<\left(\delta z\right)\right>^2,
\end{equation}
and is found to be $\sigma_z=0.039$. 
The standard deviation for the redshift range $0<z_{\text{phot}}<0.4$, within which we choose to work, 
is $\sigma_z=0.035$.

In Fig. ~\ref{fig:zspec_vs_zphot} we compare our photo-$z$ estimates with the publicly available photo-$z$ from the SDSS website (\citealp{Oyaizu2008}, tables \verb+photoz1+ and \verb+photoz2+). For this comparison we plot the redshift error as a function of photo-$z$. We then calculate the mean and the standard deviation of $\delta z$ for photo-$z$ bins of width $\Delta z_\text{{phot}} = 0.05$. The number of catastrophic outliers (galaxies with $|\zphot-\zspec|>3\sigma_z$) for the GAMA calibrated photo-$z$ is 1 percent or less for all photo-$z$ bins. We work in fixed photo-$z$ bins, because all our derived quantities are based on the photo-$z$ estimates.
This way, any biases with estimated photo-$z$ are readily apparent.
Our results based on the GAMA training set outperform the SDSS results --- for the redshift range $0.01<z_{\text{phot}}<0.4$, we obtain essentially unbiased redshift estimates, given the observed scatter. 
The scatter, in turn, increases with redshift. 
We note, however, that the \verb+photoz2+ catalogue from SDSS DR7 has been improved with the addition of $p(z)$ estimates which are designed to perform much better in recovering the total redshift probability distribution function of all galaxies \citep{Cunha2009}. Since it is still not clear how to directly relate a redshift pdf to absolute magnitude and colour for a given galaxy, our approach for the study of luminosity- and colour-dependent clustering is easier to interpret. 

In Appendix~\ref{clust:crosscorr}, we quantify the photo-$z$ error and possible contamination between redshift bins by cross-correlating photo-$z$ bins which are more than $2\sigma_{z}$ apart. We find, as expected, that the residual cross-correlation of the different photo-$z$ bins is negligible compared to their auto-correlation. 

The distribution of photo-$z$ errors is in general non-Gaussian, albeit less pronounced in the case of a complete training set. Photo-$z$ errors also propagate asymmetrically in absolute magnitude: for a given redshift error, the error induced in absolute magnitude is larger at low-$z$ and smaller at high-$z$, and thus a photo-$z$ analysis is more tolerant to redshift errors for objects at high-$z$.
For that reason, it is common practice to scale the redshift error by the quantity $1/(1+z_\text{{phot}})$. 
Taking into account this redshift stretch, $\sigma_0$ can be defined as
\begin{equation}\label{eq:z_dispersion}
\sigma_0^2=\left<\left(\frac{\delta z}{1+z_{\text{phot}}}\right)^2\right>- \left<\left(\frac{\delta z}{1+z_{\text{phot}}}\right)\right>^2,
\end{equation}
giving $\sigma_0=0.032$. 

We exclude from our analysis galaxies with $\zphot < 0.002$ or $\zphot > 0.4$. ANNz provides a photo-$z$ error calculated from the photometric errors. Using our test set, we find that this error underestimates the true photo-$z$ error (given from equation \ref{eq:errors}). We therefore apply a cut on the output parameter $z_{\text{err}}$ of ANNz at $z_{\text{err}} < 0.05$. These cuts eliminate $\sim4$ per cent of the galaxies. Cross-checks show that the correlation function measurements do not change if we use a less strict cut, but the chosen cut does improve the $N(z)$ estimates. The final number of galaxies after this cut is 4,289,223. We summarize the changes in the number of galaxies in our sample in Table~\ref{table:number_cuts}. We use Petrosian magnitudes to divide galaxies by luminosity and model magnitudes to calculate galaxy colours.

\begin{table}
\caption{The change in the total number of galaxies as a result of the cuts applied in various stages of the analysis. }
\label{table:number_cuts}
\begin{center}
\begin{math} 
\begin{array}{lc}
\hline\hline
\text{Cut description} & \text{Number of galaxies left} \\
\hline
\text{None} & 4,914,434 \\
\text{Colour cuts (Table~\ref{table:colour_cuts}}) & 4,890,965 \\
\text{Masking} & 4,511,011 \\
\text{$z^{(\text{ANNz})}_{\text{err}}<0.05$ \& $0.002<\zphot<0.4$}  & 4,289,223 \\
\hline
\end{array}
\end{math}
\end{center}
\end{table}

The photo-$z$ work presented here is similar, but not identical, to that of \citet{Parkinson2012}. The latter is appropriate for even fainter SDSS magnitudes as it uses, in its training and validation, all GAMA galaxies with $r_{petro} < 19.8$ and fainter zCOSMOS galaxies \citep{Lilly2007} matched to SDSS DR7 imaging. Minor differences in the two photo-$z$ pipelines, such as the inclusion of different light profile measurements, do not significantly affect the estimated photo-$z$, which present a similar scatter around the underlying spectroscopic distribution.  Our photo-$z$ agree with those of \citet{Parkinson2012} within the estimated errors.

\subsection{Division by redshift, absolute magnitude and colour}\label{clust:bins}
\begin{figure}
\begin{center}
\includegraphics[width=\linewidth]{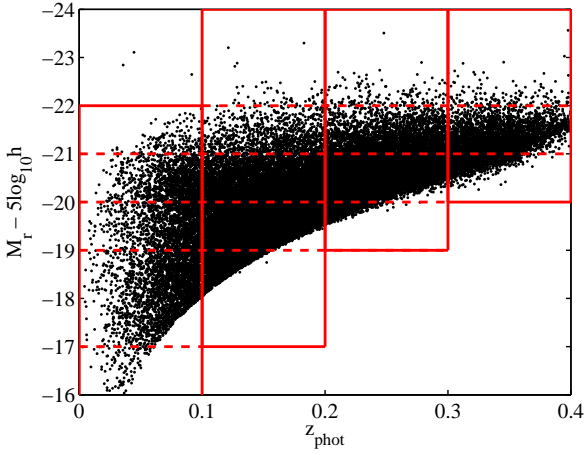}
\caption{$r$-band absolute magnitude against photo-$z$ for our photometric sample. Solid red lines show the boundaries of our samples in photo-$z$ and absolute magnitude and dashed lines the further split in absolute magnitude bins. Only 1 percent of the galaxies are shown. }
\label{fig:zphot_absmagr}
\end{center}
\end{figure}

\begin{figure}
\begin{center}
\includegraphics[width=1\linewidth]{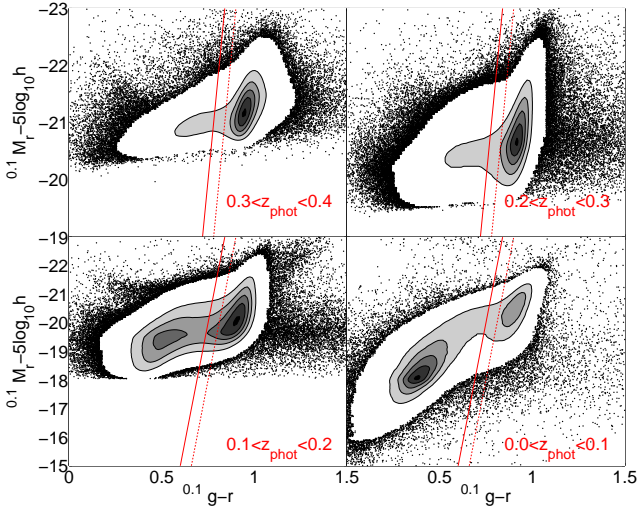}
\caption{$r$-band absolute magnitude against $^{0.1}(g-r)$ colour (both $k$-corrected and passively evolved to $z=0.1$) for galaxies split in photo-$z$ bins. Solid red lines show the colour cut for red and blue populations suggested by \citet{Loveday2012} and used in this work, while dashed red lines the colour cut used by \citet{Zehavi2010}. }
\label{fig:colgr_absr}
\end{center}
\end{figure}

\begin{figure}
\begin{center}
\includegraphics[width=1\linewidth]{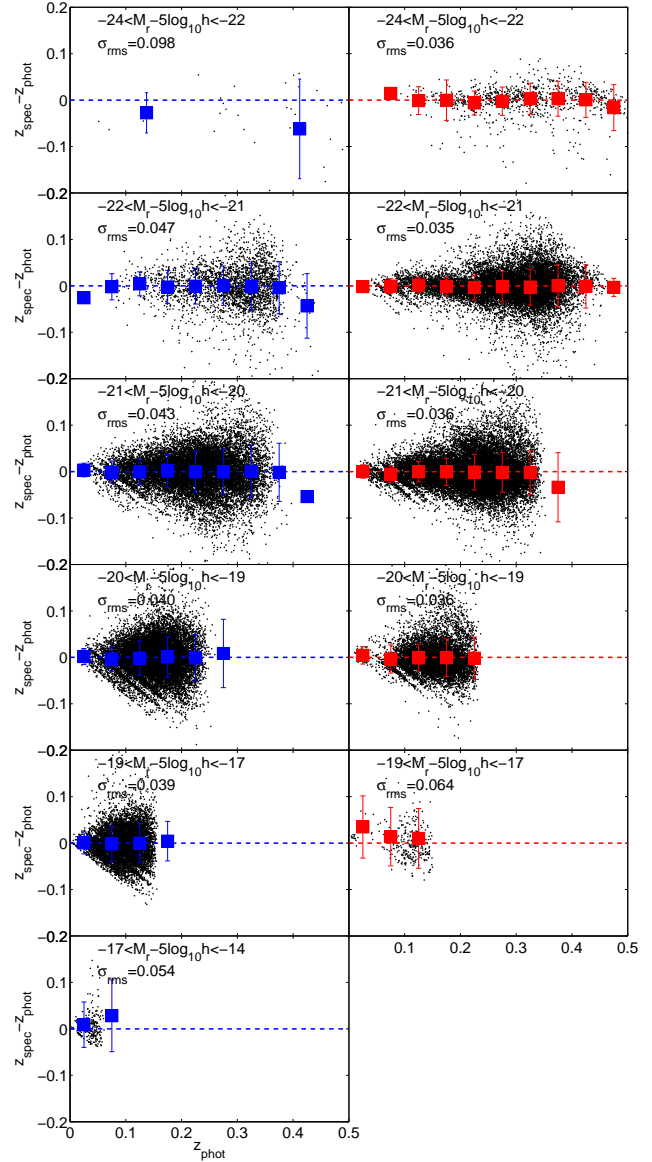}
\caption{Redshift error against photo-$z$ for our luminosity and colour-selected GAMA subsamples. The mean redshift error and standard deviation in bins of photo-$z$ are shown by the coloured squares and error bars, while the root mean square standard deviation, $\sigma_{\rm rms}$, is listed in each panel. The faint red sample has been omitted due to the small number of galaxies that it contains. } 
\label{fig:zphot_comp_samp}
\end{center}
\end{figure}

Galaxy magnitudes are $k+e$-corrected to $\zphot=0.1$, using {\sc kcorrect} version 4.1.4 \citep{Blanton2007} and the passive evolution parameter $Q=1.62$ of \cite{Blanton2003}. In this simple model, the evolution-corrected absolute magnitude is given by $M_{\rm corr} = M - Q(z - z_0)$, where $z_0 = 0.1$ is the reference redshift.
We note that \cite{Loveday2012} using GAMA found $Q=0.7$, which would change evolution-corrected magnitudes by $\approx 0.3$ mag at $z=0.4$. Approximately equal deviations in absolute magnitude will be induced in our high-$z$ blue galaxy samples, if we use a colour-dependent Q (e.g. \citealp{Loveday2012}).
Assuming a global value for $Q$ however allows for a more direct comparison with the SDSS-based clustering studies of Zehavi et al. (2005, 2011). Galaxy colours, derived from SDSS model magnitudes, are referred to as $^{0.1}(g-r)$, while absolute magnitude are derived using the $r$-band Petrosian magnitude (to match the GAMA redshift survey selection). Fig.~\ref{fig:zphot_absmagr} shows that the $r$-band absolute magnitude extends to $M_r - 5 \log h = -16$ mag with a few galaxies reaching as faint as $M_r - 5 \log h = -14$ mag. 

We split our galaxy sample in photo-$z$ as well as luminosity bins. 
%Had we used spectroscopic redshifts, these splits would have created a series of volume-limited samples. However, as we show in Section \ref{clust:dndz}, this is no longer the case for samples selected according to photo-$z$ and photo-$z$-derived quantities. 
Our samples are shown in Fig~\ref{fig:zphot_absmagr}. Initially we define four photo-$z$ bins in the redshift range $0<\zphot<0.4$ and then we further split each photo-$z$-defined sample into six absolute magnitude bins in the range $-24<M_r - 5 \log h<-14$. Thus our photo-$z$ catalogue offers the opportunity for a clustering analysis over the luminosity range $0.03L^*\lesssim L\lesssim 8L^*$, spanning almost three orders of magnitude in $L/L^*$. 

In Fig~\ref{fig:zphot_absmagr} some of these redshift-magnitude bins extending beyond the survey flux limit are only partially occupied by galaxies in terms of photometric redshifts and photo-$z$ derived absolute magnitudes.  The true redshift and absolute magnitude distributions for each bin are recovered by Monte-Carlo resampling, as discussed in Section 4.3.

Fig.~\ref{fig:colgr_absr} shows colour-magnitude diagrams for our sample split in photo-$z$ bins. The colour bimodality is evident at $^{0.1}(g-r) \simeq 0.8$ for all photo-$z$ bins. We have adopted the tilted colour cuts defined by \citet{Loveday2012},
\begin{equation}\label{eq:colour_cut}
M_r - 5 \log h = 5 - 33.3\times ^{0.1}(g-r)_{\text{model}},
\end{equation}
which is a slightly modified version of the colour cut used by \citet{Zehavi2010}, also shown in Fig.~\ref{fig:colgr_absr}. 

In Fig.~\ref{fig:zphot_comp_samp} we plot the photo-$z$ error against photo-$z$ for galaxies subdivided into subsamples, where we again have used
{\em photometric} redshifts to estimate galaxy luminosities and colours.
There are no obvious systematic biases of
$\zspec - \zphot$ for any of the subsamples, although we do note that the
most luminous (faintest) bin contains very few blue (red) galaxies. 

The relatively good photo-$z$s notwithstanding, our analysis does not eliminate completely the main systematic error of neural network derived photo-$z$, which is the overestimation of low redshifts and the underestimation of high redshifts \citep[see e.g. Fig. 7 of ][]{megaZ}. As a result, a number of faint galaxies have their redshift overestimated and hence appear brighter in our sample. We note that there is a discrepancy between the fraction of faint red objects in the luminosity bin $-19<M_r - 5 \log h<-17$ between this work and \cite{Zehavi2010}, which is most probably caused by this systematic shift (see Table \ref{table:r0}). It is possible to cure this by Monte-Carlo resampling the photo-$z$s with their respective errors and then rederive the absolute magnitudes and colours, but we do not pursue this here.   

\subsection{Photometric redshift distribution(s)}\label{clust:dndz}

\begin{figure}
\begin{center}
\includegraphics[width=1.\linewidth]{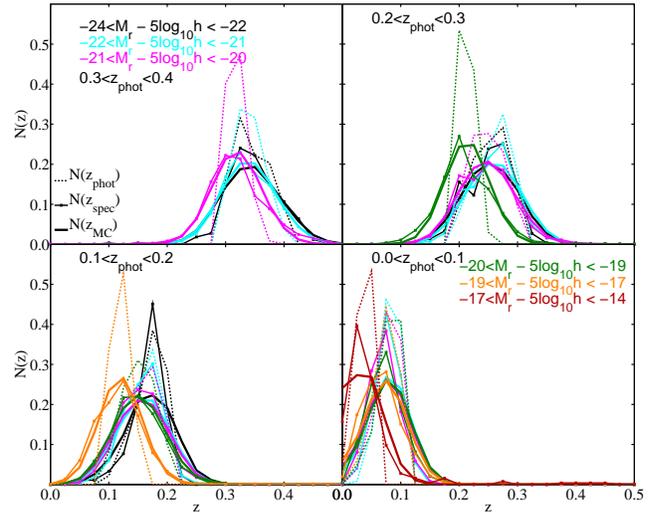}
\caption{Estimates of the underlying redshift distribution for the luminosity samples used in the clustering analysis. Thin solid lines show the photo-$z$ distribution, which is the basis for the selection, dotted lines the true spectroscopic redshift distribution from GAMA and thick solid line the average distribution inferred from 100 Monte-Carlo resamplings of the photo-$z$ distribution using equation \ref{eq:noise}.  
}
\label{fig:dndz_all}
\end{center}
\end{figure}

\begin{figure}
\begin{center}
\includegraphics[width=1.\linewidth]{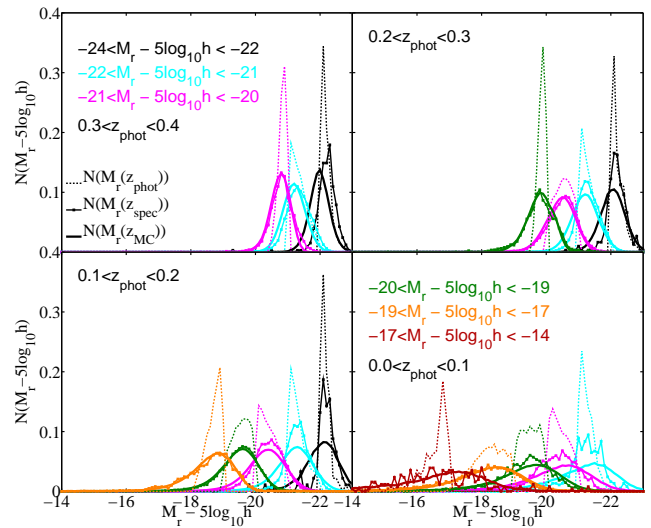}
\caption{The r-band absolute magnitude distribution for GAMA galaxies with $\rpet<19.4$ split into photo-$z$ and  photo-$z$-derived absolute magnitude slices. Magnitude distributions shown by dashed lines are derived from the raw photo-$z$, by thin lines from the underlying spectroscopic redshifts and by thick lines from the Monte-Carlo derived magnitudes.  The latter reproduces the true underlying spec-$z$ inferred magnitude distribution rather well; however for a few samples there is a discrepancy between the spec-$z$-derived and the Monte-Carlo-derived distributions. All MC absolute magnitude estimates are $K$-corrected and passively evolved following the procedure described in Section~\ref{clust:bins}. 
}
\label{fig:mr_distr}
\end{center}
\end{figure}

Despite the fact that ANNz gives fairly accurate and unbiased photo-$z$s for calculations in broad absolute magnitude bins or photo-$z$ bins, in order to translate the two dimensional clustering signal to the three dimensional one using equation~\ref{equ:limber_inv}, the underlying true $dN/dz$ is needed. In this work we loosely follow the approach given in 
\citet{Parkinson2012}, \citep[see also][]{Driver2010}. The GAMA spectroscopic sample is highly representative and it allows us to calculate the true redshift errors as a function of photo-$z$ for all objects in GAMA with $\rpet<19.4$. Then, under the assumption of a Gaussian photometric error distribution in each photo-$z$ bin, we perform a Monte-Carlo resampling of the ANNz predictions for photo-$z$s. This is equivalent to replacing each photo-$z$ derived from ANNz with the quantity $z_{\text{MC}}$ drawn from a Gaussian distribution, using a photo-$z$ dependent standard deviation, $\sigma(z_{\text{phot}}^{\text{(bin)}}) = \delta z_{\text{phot}}^{\text{(bin)}}$:
\begin{equation}\label{eq:noise}
z_{\text{MC}} = G\lbrack \mu=\zphot, \sigma = \sigma_{\text{phot}}(1+\zphot)\rbrack.
\end{equation}  
Note that \emph{convolving} the imprecise photo-$z$ with additional scatter improves the $N(z)$ redshift distribution: in other words the photo-$z$ process \emph{deconvolves} the $N(z)$ and makes it artificially
narrow.

All our sample selections in Fig.~\ref{fig:dndz_all} have been made using the photo-$z$ derived absolute magnitude $M_r-5\text{log}h$. We then use the accurate spectroscopic information from GAMA to assess how well Monte-Carlo resampling compares to the underlying true $dN/dz$. 
Since the GAMA area is much smaller than the SDSS area, we do not wish to recover the exact spectroscopic redshift distribution, merely to match a smoothed version thereof. Our test shows that MC resampling performs rather well in recovering the true $dN/dz$. This method performs even better with a larger number of objects, which indicates that we are still dominated by statistical errors and therefore there is room for improvement in future when larger spectroscopic training sets will be available.   
Nevertheless, as an incorrect redshift distribution can cause a systematic error in $r_0$, in Appendix~\ref{clust:nz} we test the sensitivity of our results to the assumed $dN/dz$, and compare results using the Monte-Carlo recovered $dN/dz$  with those from the weighting method proposed by \citet{Cunha2009}.

Fig.~\ref{fig:mr_distr} shows, for all samples split by photo-$z$ and photo-$z$-derived absolute magnitude, the photo-$z$-derived, the true underlying and the Monte-Carlo inferred absolute magnitude distributions (as dashed, thin and thick solid lines respectively). We note that the photo-$z$ derived absolute magnitude estimates in Fig. ~\ref{fig:mr_distr} are obtained from the resampled redshifts and not by resampling the absolute magnitudes per se. We then $k+e$-correct every Monte-Carlo absolute magnitude realization using the procedure described in Section~\ref{clust:bins}. As expected, the true underlying distribution extends well beyond the photo-$z$ inferred luminosity bins, but is yet again rather well described by the Monte-Carlo inferred distribution. 

It is crucial that we have a good understanding of the true underlying absolute magnitude for all our samples. For galaxy clustering studies with spectroscopic redshifts it is desirable to work with volume-limited samples.  Using photometric redshifts, however, one can form only approximately volume-limited samples, since photo-$z$ uncertainties will propagate into absolute magnitude estimates.  Essentially, any tophat absolute magnitude distribution, as selected using photo-$z$, corresponds to a wider true absolute magnitude distribution, as shown in Fig. 7. This is rather similar to selecting galaxies from a photometric redshift bin and then convolving the initial tophat distribution with the photo-$z$ error distribution in order to obtain the true $N(z)$.     
However, using the $w(\theta)$ statistic and an accurate $dN/dz$ for that particular galaxy sample we can extract its respective spatial clustering signal, which would then correspond to the $z_{\text{MC}}$ derived absolute magnitude. Direct comparisons with other studies can then be made, modulo the extent of the overlap between the two absolute magnitude distributions.

\section{Results for the two-point correlation function}\label{clust:corres}

\subsection{Luminosity and redshift dependence}\label{clust:lum_dependence}
\begin{figure*}
\begin{center}
\includegraphics[width=1.\linewidth]{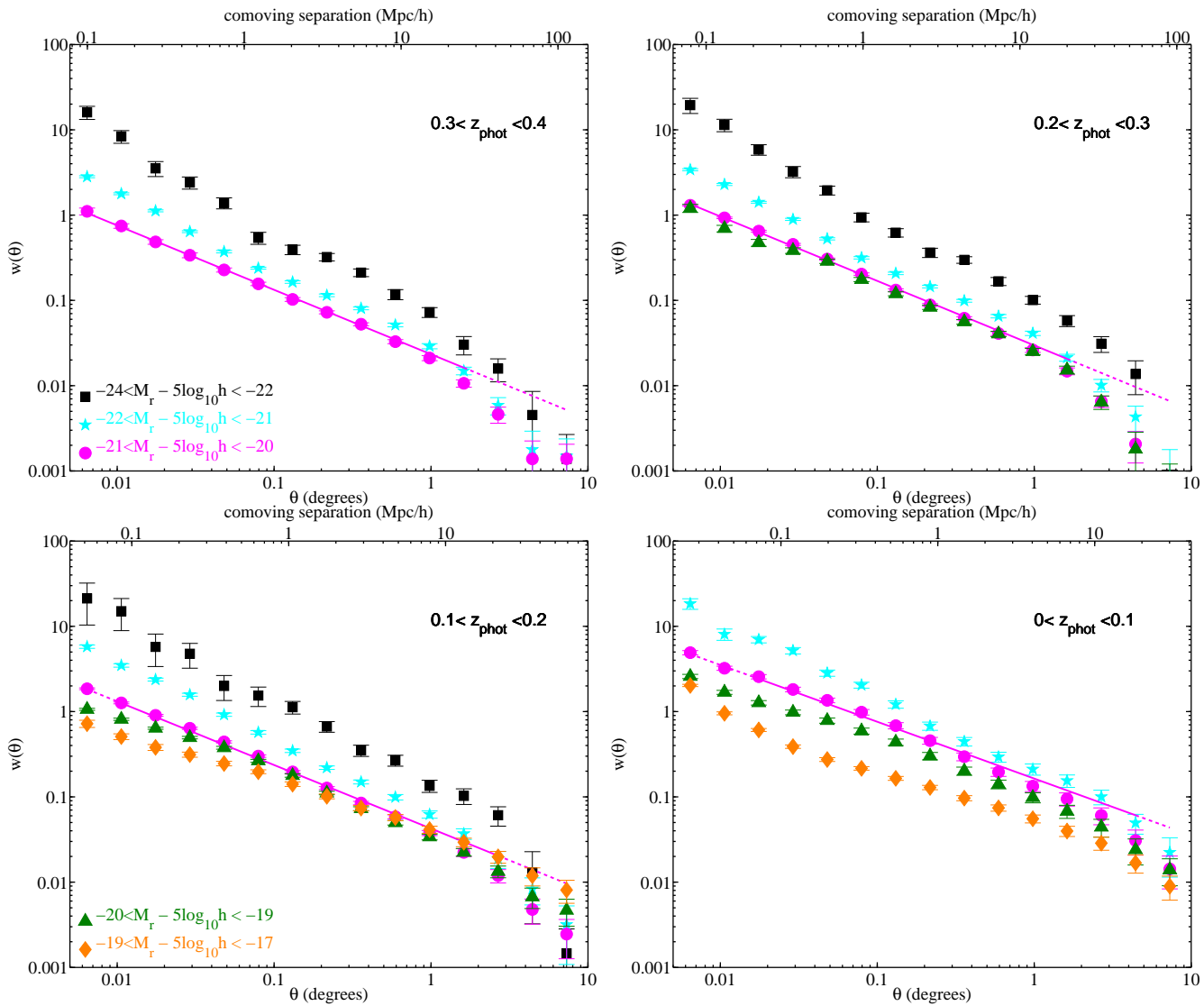}
\caption{Two-point angular correlation functions $w(\theta)$ of our samples split into photo-$z$ bins and six photo-$z$-inferred absolute magnitude bins, as indicated in each panel, with jackknife errors. The solid lines show power law fits estimated using the full covariance matrix for the $L^*$ sample. Dotted lines show the extension of the power law fits on scales $<0.1 h^{-1}\text{Mpc}$ and $>20 h^{-1}\text{Mpc}$. }
\label{fig:cor_absmagr}
\end{center}
\end{figure*}

\begin{figure*}
\begin{center}
\includegraphics[width=\linewidth]{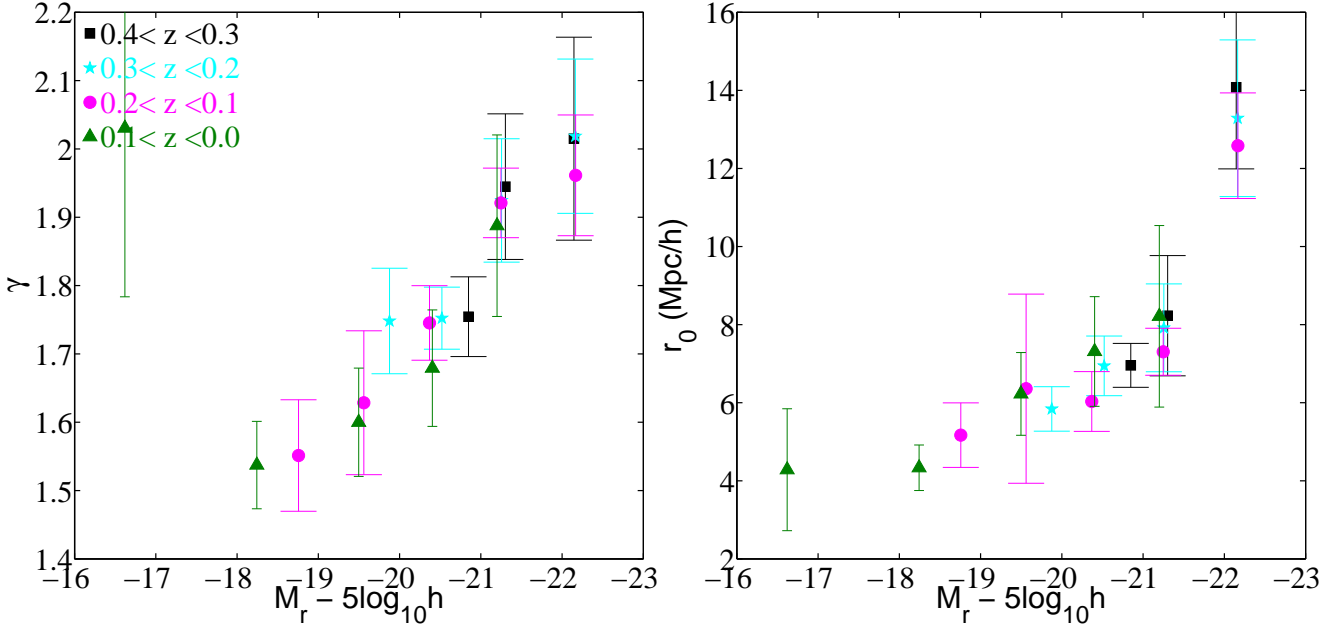}
\caption{Left: Power law slope, $\gamma$, as a function of absolute magnitude and redshift. Right: Real space correlation length, $r_{0}$, as a function of absolute magnitude and redshift. Absolute magnitude ranges for which $r_0$ and $\gamma$ measurements are valid are given in Table \ref{table:r0}.
} 
\label{fig:absmagr_gamma_r0}
\label{r0_gamma}
\end{center}
\end{figure*}

\begin{table*}
\caption{Clustering properties of luminosity-selected samples. Col. 1 lists the photo-$z$ based absolute magnitude ranges, col. 2 the median absolute magnitude and the associated 16$^{\rm th}$ and 84$^{\rm th}$ percentiles from the Monte-Carlo resampling (Fig.~\ref{fig:mr_distr}) and col. 3 the number of galaxies in each sample. Cols. 4, 5 and 6 list respectively the slope, $\gamma$, the correlation length, $r_{0}$, and the reduced $\chi^2$, $\chi^2_{\nu}$, of the power law fit as defined in Section~\ref{clust:xi}. Cols. 7, 8 and 9 show the same information but for power law fits using only the diagonal elements of the covariance matrix. All power law fits are approximately over the comoving scales $0.1<r<20 \ h^{-1}$~Mpc.  
Finally col. 10 presents the relative bias at 5 $h^{-1}$~Mpc measured using equation~\ref{eq:relbias}.
}
\label{table:r0} \vspace{3mm}
\begin{math}
\begin{array}{rrrrrrrrrrr}
%\begin{array}{cccccccccccc}
  \hline\hline
  \multicolumn{1}{c} {\mbox{Sample}} & 
    \multicolumn{1}{c}  {\mbox{Magnitude}^{\text{(MC)}}} & 
        %\multicolumn{1}{c} {z_{\text{med}}} & 
          \multicolumn{1}{c} {N_{\rm gal}} & 
            \multicolumn{1}{c} {\gamma} & 
              \multicolumn{1}{c} {r_{0}} & 
                \multicolumn{1}{c} {\chi^2_{\nu}} & 
                  \multicolumn{1}{c} {\gamma^{(d)}} & 
                    \multicolumn{1}{c} {r^{(d)}_{0}} & 
                      \multicolumn{1}{c} {{\chi^{(d)}}^2_{\nu}} & 
                       %\multicolumn{1}{c} {\mbox{Scale}} & 
	  	      \multicolumn{1}{c} {b/b^{*}} \\
  \multicolumn{1}{c}{M_r - 5 \log h} & 
    \multicolumn{1}{c} {M_r - 5 \log h} & & &   \multicolumn{1}{c} {[h^{-1} \text{Mpc}]} & & &   \multicolumn{1}{c} {[h^{-1} \text{Mpc}]} &  &   \multicolumn{1}{c} {[h^{-1} \text{Mpc}]} & \\
\hline
\multicolumn{10}{c}{\mbox{All colours} \quad 0.3<\zphot<0.4}\\
\hline
\lbrack   -24,  -22 ) &    -22.0^{ -0.2}_{+  0.2} &      13257 &     2.01 \pm  0.15 &    14.08 \pm  2.09 &  3.41 &     2.02 \pm  0.09 &    13.68 \pm  1.22 &   2.6 &  2.13 \pm  0.30 \\
\lbrack   -22,  -21 ) &    -21.2^{ -0.3}_{+  0.3} &     339834 &     1.94 \pm  0.11 &     8.23 \pm  1.54 & 28.08 &     1.91 \pm  0.09 &     8.46 \pm  1.06 &  13.0 &  1.22 \pm  0.22 \\
\lbrack   -21,  -20 ) &    -20.8^{ -0.2}_{+  0.2} &     158860 &     1.75 \pm  0.06 &     6.96 \pm  0.56 &  3.76 &     1.78 \pm  0.05 &     6.80 \pm  0.33 &   1.8 &  1.00 \pm  0.01 \\
\hline
\multicolumn{10}{c}{\mbox{All colours} \quad 0.2<\zphot<0.3}\\
\hline
\lbrack   -24,  -22 ) &    -22.0^{ -0.3}_{+  0.3} &      12294 &     2.02 \pm  0.11 &    13.29 \pm  2.01 &  2.37 &     2.01 \pm  0.07 &    13.17 \pm  1.13 &   1.7 &  2.02 \pm  0.32 \\
\lbrack   -22,  -21 ) &    -21.2^{ -0.4}_{+  0.3} &     284969 &     1.92 \pm  0.09 &     7.92 \pm  1.13 & 10.91 &     1.90 \pm  0.06 &     8.12 \pm  0.70 &   5.5 &  1.17 \pm  0.17 \\
\lbrack   -21,  -20 ) &    -20.4^{ -0.3}_{+  0.4} &     930539 &     1.75 \pm  0.05 &     6.94 \pm  0.76 &  7.96 &     1.77 \pm  0.05 &     6.74 \pm  0.36 &   3.3 &  1.00 \pm  0.03 \\
\lbrack   -20,  -19 ) &    -19.8^{ -0.3}_{+  0.3} &     122870 &     1.75 \pm  0.08 &     5.84 \pm  0.57 &  2.44 &     1.76 \pm  0.06 &     5.84 \pm  0.29 &   1.5 &  0.86 \pm  0.10 \\
\hline
\multicolumn{10}{c}{\mbox{All colours} \quad 0.1<\zphot<0.2}\\
\hline
\lbrack   -24,  -22 ) &    -22.0^{ -0.4}_{+  0.3} &       4311 &     1.96 \pm  0.09 &    12.58 \pm  1.35 &  0.59 &     1.95 \pm  0.08 &    12.57 \pm  1.13 &   0.4 &  2.10 \pm  0.35 \\
\lbrack   -22,  -21 ) &    -21.2^{ -0.4}_{+  0.5} &     106728 &     1.92 \pm  0.05 &     7.31 \pm  0.60 &  3.56 &     1.92 \pm  0.04 &     7.40 \pm  0.32 &   1.7 &  1.22 \pm  0.18 \\
\lbrack   -21,  -20 ) &    -20.3^{ -0.5}_{+  0.5} &     604181 &     1.75 \pm  0.05 &     6.03 \pm  0.77 &  7.16 &     1.78 \pm  0.06 &     5.85 \pm  0.43 &   3.9 &  1.00 \pm  0.05 \\
\lbrack   -20,  -19 ) &    -19.5^{ -0.4}_{+  0.5} &     916563 &     1.63 \pm  0.11 &     6.36 \pm  2.42 & 42.40 &     1.71 \pm  0.10 &     5.81 \pm  0.75 &  11.7 &  1.03 \pm  0.30 \\
\lbrack   -19,  -17 ) &    -18.6^{ -0.4}_{+  0.6} &     211336 &     1.55 \pm  0.08 &     5.17 \pm  0.83 &  4.41 &     1.58 \pm  0.07 &     4.89 \pm  0.34 &   1.6 &  0.87 \pm  0.16 \\
\hline
\multicolumn{10}{c}{\mbox{All colours} \quad 0.0<\zphot<0.1}\\
\hline
\lbrack   -22,  -21 ) &    -21.1^{ -0.7}_{+  0.8} &      19218 &     1.89 \pm  0.13 &     8.21 \pm  2.32 &  6.36 &     1.88 \pm  0.07 &     8.09 \pm  0.80 &   1.6 &  1.15 \pm  0.43 \\
\lbrack   -21,  -20 ) &    -20.3^{ -0.7}_{+  0.9} &     122787 &     1.68 \pm  0.09 &     7.31 \pm  1.40 &  9.00 &     1.75 \pm  0.05 &     6.84 \pm  0.50 &   2.1 &  0.99 \pm  0.23 \\
\lbrack   -20,  -19 ) &    -19.4^{ -0.6}_{+  0.8} &     155147 &     1.60 \pm  0.08 &     6.23 \pm  1.06 &  9.08 &     1.65 \pm  0.08 &     6.10 \pm  0.64 &   4.5 &  0.86 \pm  0.20 \\
\lbrack   -19,  -17 ) &    -18.1^{ -0.8}_{+  1.0} &     271389 &     1.54 \pm  0.06 &     4.33 \pm  0.58 &  6.20 &     1.58 \pm  0.09 &     3.97 \pm  0.24 &   2.9 &  0.65 \pm  0.18 \\
\lbrack   -17,  -14 ) &    -16.6^{ -0.9}_{+  1.4} &      14659 &     2.03 \pm  0.25 &     4.28 \pm  1.56 &  5.82 &     2.00 \pm  0.28 &     4.41 \pm  1.03 &   2.1 &  0.62 \pm  0.25 \\
\hline  
\end{array}
\end{math}
\end{table*}

We first calculate the angular correlation function $w(\theta)$ for our samples selected on absolute magnitude and photometric redshift over angular scales from 0.005 to 9.4 degrees, in 15 equally spaced bins in 
log($\theta$)\footnote{Initially our analysis was done down to $\theta = 0.001$ degrees. However, as shown in Section \ref{clust:faintblue} and Appendix \ref{clust:systematics_faint}, the data is not reliable enough on such small scales.}. In a flux-limited survey like SDSS, intrinsically bright galaxies dominate at high redshifts and intrinsically faint objects dominate at low redshifts (see Fig.~\ref{fig:colgr_absr}). For that reason, we calculate $w(\theta)$ for the 17 well-populated samples given in Table \ref{table:r0}. Errors are estimated using the jackknife technique,
with the covariance matrix given by equation~\ref{eq:cov}. Even if the
validity of a given error method based on data alone is still widely
debated, it is commonly accepted that the jackknife method is adequate
for angular clustering studies \citep[see e.g.][]{Cabre2007}, while
for 3-D clustering measurements, \cite{Norberg2009} have shown that
the jackknife method suffers from some limitations, in particular on small scales.

Our angular correlation function measurements are broad and probe both highly non-linear and quasi-linear scales. Fig.~\ref{fig:cor_absmagr} presents galaxy angular correlation functions for six photo-$z$ selected absolute magnitude bins. We show the angular scale (lower $x$-axis), used for the correlation function estimation, and the corresponding comoving scale estimated at the mean redshift of the sample (upper $x$-axis). 

Over the range of angular scales fitted, chosen to correspond to approximately 0.1--20 $h^{-1}$~Mpc 
comoving separation according to the mean redshift of each sample,
the angular correlation function can be reasonably well approximated by a power law, equation~\ref{equ:powlaw_w}. We perform power law fits, both with the full covariance matrix and with the diagonal elements only. The power law fits for our $L^*$ sample are shown in Fig.~\ref{fig:cor_absmagr}. Dotted lines in Fig.~\ref{fig:cor_absmagr} show the extension of the power laws beyond the scales over which they were fitted. The resulting correlation lengths, $r_{0}$, slopes, $\gamma$, and quality of the fits as given by the reduced $\chi^2$, $\chi^2_{\nu}$, for all samples are listed in Table~\ref{table:r0}. 

The luminosity dependence of galaxy clustering is present in all photo-$z$ shells: the shape and the amplitude of the angular correlation function differ for galaxies with different luminosity. The amplitude of the angular correlation function decreases as we go from bright to faint galaxies for all photo-$z$ bins. The slope of the correlation function also decreases with decreasing luminosity, very much in line with the change in the fraction of red and blue galaxies. As observed in Section~\ref{clust:colour_dependence}, red (blue) galaxies dominate the brightest (faintest) luminosity bins, with red galaxies preferentially having a steeper correlation function slope than blue galaxies. 
  
For each sample, we estimate the correlation length $r_0$ via equation~\ref{equ:limber_inv} using the Monte-Carlo inferred redshift distribution described in Section~\ref{clust:dndz}. The redshift distribution $dN/dz$ is calculated separately for each sample, as shown in Fig~\ref{fig:dndz_all}. 
%Unexpectedly, we find systematic differences in the correlation lengths which depend on the exact $dN/dz$ used.  %JL: why is this unexpected?
In Appendix~\ref{clust:nz} we investigate the effects of the assumed $dN/dz$ on the recovered correlation length $r_0$, and show that the adopted $dN/dz$ recovery method compares favourably with the true underlying $dN/dz$, as obtained from the smoothed $dN/dz_{\text{spec}}$.

For our luminosity bins in the redshift range $0<z<0.1$, the correlation length is found to decrease as we go to fainter absolute magnitudes, from $8.21\pm2.32 \ h^{-1}\text{Mpc}$ ($-22<M_r -5 \log h<-21$) to  $4.28 \pm1.56 h^{-1}\text{Mpc}$ ($-19<M_r -5 \log h<-17$). This is very much in line with the recent results of \cite{Zehavi2010}. Moreover, we do not observe strong evolution with redshift for samples of fixed luminosity. All $r_0$ and $\gamma$ measurements are shown in Fig.~\ref{r0_gamma}.

There are two main sources of error in the $r_0$ estimates: (a) the correlated uncertainties on the power law parameters $\gamma$ and $A_w$ which propagate through equation~\ref{equ:limber_inv} to $r_0$; (b) statistical and systematic uncertainties in the modelling of the underlying redshift distribution. The $w(\theta)$ uncertainties and the induced error on $r_0$ and $\gamma$ are obtained using the standard deviation from the distribution of JK resampling estimates (Section \ref{clust:pix}). As in the case of the covariance matrix, these uncertainties are multiplied by a factor of $N-1$ \citep{Norberg2009}. The $dN/dz$ uncertainties are investigated in great detail in Appendix~\ref{clust:nz}, where we show that the Monte-Carlo inferred $dN/dz$ performs best, while still returning a residual systematic uncertainty of $\pm 0.2 h^{-1}\text{Mpc}$ on $r_0$ that depends on the sample considered. We find that both sources of uncertainty have a comparable contribution to the errors. In Table~\ref{table:r0} we quote the total error on the correlation length after adding the two (independent) errors in quadrature.

\subsection{Luminosity, redshift and colour dependence}\label{clust:colour_dependence}

\begin{figure*}
\begin{center}
\includegraphics[width=1.\linewidth]{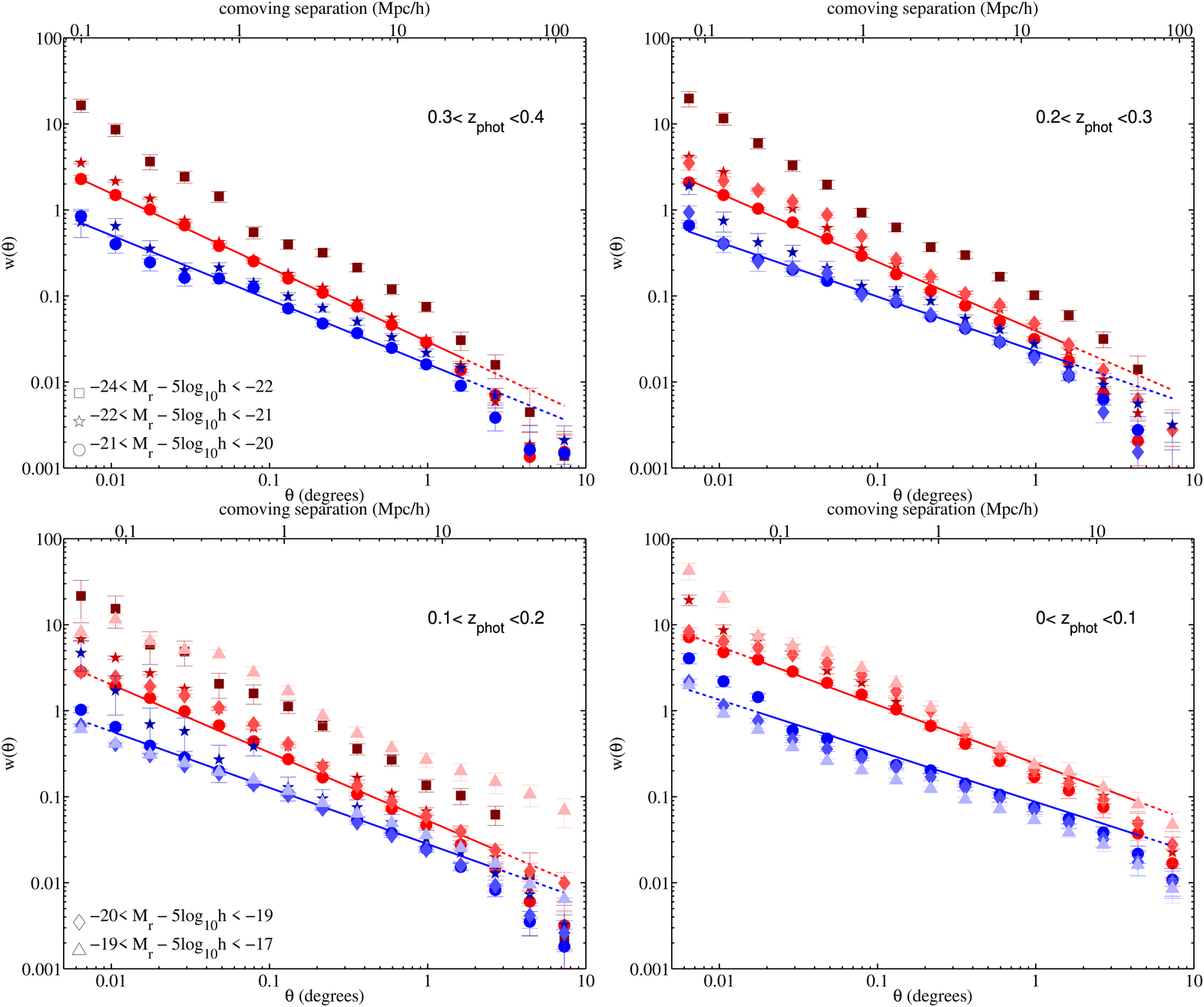}
\caption{Two-point angular correlation functions $w(\theta)$ split by absolute magnitude and colour, with red circles (blue squares) showing the red (blue) sample. Colour gradients indicate the transition from bright (darker shade) to faint (lighter shade) luminosities. Lines are as in Fig.~\ref{fig:cor_absmagr}. The faintest (brightest) sample does not contain enough red (blue) galaxies to robustly estimate $w(\theta)$.}
\label{fig:cor_colours}
\end{center}
\end{figure*}

\begin{table*}
\caption{Clustering properties of luminosity-selected red galaxies. Columns are the same as in Table \ref{table:r0}.
}
\label{table:r02} \vspace{3mm}
\begin{math}
\begin{array}{rrrrrrrrrrr}
%\begin{array}{cccccccccccc}
  \hline\hline
  \multicolumn{1}{c} {\mbox{Sample}} & 
    \multicolumn{1}{c}  {\mbox{Magnitude}^{\text{(MC)}}} & 
        %\multicolumn{1}{c} {z_{\text{med}}} & 
          \multicolumn{1}{c} {N_{\rm gal}} & 
            \multicolumn{1}{c} {\gamma} & 
              \multicolumn{1}{c} {r_{0}} & 
                \multicolumn{1}{c} {\chi^2_{\nu}} & 
                  \multicolumn{1}{c} {\gamma^{(d)}} & 
                    \multicolumn{1}{c} {r^{(d)}_{0}} & 
                      \multicolumn{1}{c} {{\chi^{(d)}}^2_{\nu}} & 
                       %\multicolumn{1}{c} {\mbox{Scale}} & 
	  	      \multicolumn{1}{c} {b/b^{*}} \\
  \multicolumn{1}{c}{M_r - 5 \log h} & 
    \multicolumn{1}{c} {M_r - 5 \log h} & & &   \multicolumn{1}{c} {[h^{-1} \text{Mpc}]} & & &   \multicolumn{1}{c} {[h^{-1} \text{Mpc}]} &  &   \multicolumn{1}{c} {[h^{-1} \text{Mpc}]} & \\
\hline
\multicolumn{10}{c}{\mbox{Red} \quad 0.3<\zphot<0.4}\\
\hline
\lbrack   -24,  -22 ) &    -22.0^{ -0.2}_{+  0.2} &      13095 &     2.02 \pm  0.15 &    13.91 \pm  2.22 &  3.01 &     2.03 \pm  0.11 &    13.65 \pm  1.86 &   2.4 &  1.78 \pm  0.26 \\
\lbrack   -22,  -21 ) &    -21.2^{ -0.3}_{+  0.3} &     287622 &     1.98 \pm  0.10 &     8.40 \pm  1.64 & 24.60 &     1.94 \pm  0.10 &     8.71 \pm  1.17 &  13.7 &  1.06 \pm  0.20 \\
\lbrack   -21,  -20 ) &    -20.7^{ -0.2}_{+  0.2} &      79073 &     1.86 \pm  0.05 &     8.19 \pm  0.54 &  1.33 &     1.88 \pm  0.05 &     8.08 \pm  0.40 &   1.2    &  1.00 \pm  0.01 \\
\hline
\multicolumn{10}{c}{\mbox{Red} \quad 0.2<\zphot<0.3}\\
\hline
\lbrack   -24,  -22 ) &    -22.0^{ -0.3}_{+  0.3} &      12200 &     2.02 \pm  0.11 &    13.33 \pm  1.95 &  1.89 &     2.01 \pm  0.07 &    13.24 \pm  1.11 &   1.8 &  1.73 \pm  0.41 \\
\lbrack   -22,  -21 ) &    -21.2^{ -0.4}_{+  0.3} &     242452 &     1.95 \pm  0.10 &     8.26 \pm  1.31 & 11.23 &     1.92 \pm  0.06 &     8.41 \pm  0.72 &   6.0 &  1.05 \pm  0.25 \\
\lbrack   -21,  -20 ) &    -20.5^{ -0.3}_{+  0.4} &     597678 &     1.81 \pm  0.06 &     8.01 \pm  1.20 & 17.10 &     1.84 \pm  0.06 &     7.69 \pm  0.52 &   6.5 &  0.98 \pm  0.04 \\
\lbrack   -20,  -19 ) &    -19.8^{ -0.3}_{+  0.3} &      44588 &     1.95 \pm  0.09 &     8.53 \pm  1.30 &  5.59 &     1.91 \pm  0.08 &     8.57 \pm  0.43 &   2.8 &  1.07 \pm  0.21 \\
\hline
\multicolumn{10}{c}{\mbox{Red} \quad 0.1<\zphot<0.2}\\
\hline
\lbrack   -24,  -22 ) &    -22.0^{ -0.4}_{+  0.3} &       4271 &     1.96 \pm  0.08 &    12.61 \pm  1.26 &  0.47 &     1.95 \pm  0.08 &    12.57 \pm  1.13 &   0.4 &  1.87 \pm  0.48 \\
\lbrack   -22,  -21 ) &    -21.2^{ -0.4}_{+  0.5} &      93975 &     1.94 \pm  0.05 &     7.56 \pm  0.71 &  2.52 &     1.93 \pm  0.04 &     7.65 \pm  0.36 &   1.6 &  1.13 \pm  0.28 \\
\lbrack   -21,  -20 ) &    -20.3^{ -0.5}_{+  0.5} &     393344 &     1.78 \pm  0.11 &     7.07 \pm  1.81 & 17.30 &     1.84 \pm  0.08 &     6.68 \pm  0.64 &   6.3 &  1.03 \pm  0.10 \\
\lbrack   -20,  -19 ) &    -19.5^{ -0.4}_{+  0.5} &     344815 &     1.71 \pm  0.20 &     9.69 \pm  5.98 & 82.81 &     1.85 \pm  0.12 &     8.19 \pm  1.26 &  16.9 &  1.33 \pm  0.66 \\
\lbrack   -19,  -17 ) &    -18.7^{ -0.4}_{+  0.5} &      12942 &     1.86 \pm  0.18 &    17.86 \pm  4.26 &  9.69 &     1.84 \pm  0.14 &    17.72 \pm  2.88 &   4.6 &  2.46 \pm  0.83 \\
\hline
\multicolumn{10}{c}{\mbox{Red} \quad 0.0<\zphot<0.1}\\
\hline
\lbrack   -22,  -21 ) &    -21.1^{ -0.7}_{+  0.9} &      18631 &     1.90 \pm  0.14 &     8.20 \pm  2.62 &  5.97 &     1.88 \pm  0.07 &     8.14 \pm  0.78 &   1.7 &  0.96 \pm  0.47 \\
\lbrack   -21,  -20 ) &    -20.4^{ -0.7}_{+  0.9} &      83541 &     1.71 \pm  0.11 &     8.82 \pm  2.34 & 10.98 &     1.79 \pm  0.07 &     7.90 \pm  0.76 &   3.2 &  0.97 \pm  0.29 \\
\lbrack   -20,  -19 ) &    -19.5^{ -0.6}_{+  0.8} &      45541 &     1.77 \pm  0.16 &    10.41 \pm  3.89 & 19.29 &     1.85 \pm  0.14 &    10.39 \pm  1.66 &   8.1 &  1.15 \pm  0.46 \\
\lbrack   -19,  -17 ) &    -18.7^{ -0.5}_{+  0.7} &       6690 &     1.88 \pm  0.13 &    11.59 \pm  2.82 &  2.65 &     1.90 \pm  0.09 &    11.77 \pm  1.32 &   1.0 &  1.43 \pm  0.51 \\
\hline
\end{array}
\end{math}
\end{table*}

\begin{table*}
\caption{Clustering properties of luminosity-selected blue galaxies. Columns are the same as in Table \ref{table:r0}.
}
\label{table:r03} \vspace{3mm}
\begin{math}
\begin{array}{rrrrrrrrrrr}
%\begin{array}{cccccccccccc}
  \hline\hline
  \multicolumn{1}{c} {\mbox{Sample}} & 
    \multicolumn{1}{c}  {\mbox{Magnitude}^{\text{(MC)}}} & 
        %\multicolumn{1}{c} {z_{\text{med}}} & 
          \multicolumn{1}{c} {N_{\rm gal}} & 
            \multicolumn{1}{c} {\gamma} & 
              \multicolumn{1}{c} {r_{0}} & 
                \multicolumn{1}{c} {\chi^2_{\nu}} & 
                  \multicolumn{1}{c} {\gamma^{(d)}} & 
                    \multicolumn{1}{c} {r^{(d)}_{0}} & 
                      \multicolumn{1}{c} {{\chi^{(d)}}^2_{\nu}} & 
                       %\multicolumn{1}{c} {\mbox{Scale}} & 
	  	      \multicolumn{1}{c} {b/b^{*}} \\
  \multicolumn{1}{c}{M_r - 5 \log h} & 
    \multicolumn{1}{c} {M_r - 5 \log h} & & &   \multicolumn{1}{c} {[h^{-1} \text{Mpc}]} & & &   \multicolumn{1}{c} {[h^{-1} \text{Mpc}]} &  &   \multicolumn{1}{c} {[h^{-1} \text{Mpc}]} & \\
\hline
\multicolumn{10}{c}{\mbox{Blue} \quad 0.3<\zphot<0.4}\\
\hline
\lbrack   -22,  -21 ) &    -21.2^{ -0.3}_{+  0.3} &      52212 &     1.71 \pm  0.07 &     6.88 \pm  0.47 &  0.78 &     1.72 \pm  0.07 &     6.87 \pm  0.38 &   0.6 &  1.14 \pm  0.12 \\
\lbrack   -21,  -20 ) &    -20.8^{ -0.2}_{+  0.3} &      79787 &     1.75 \pm  0.06 &     5.86 \pm  0.49 &  1.52 &     1.75 \pm  0.10 &     5.83 \pm  0.44 &   1.3 &  1.00 \pm  0.01 \\
\multicolumn{10}{c}{\mbox{Blue} \quad 0.2<\zphot<0.3}\\
\hline
\lbrack   -22,  -21 ) &    -21.2^{ -0.3}_{+  0.3} &      42517 &     1.74 \pm  0.11 &     6.42 \pm  0.81 &  3.05 &     1.75 \pm  0.12 &     6.46 \pm  0.57 &   1.5 &  1.17 \pm  0.14 \\
\lbrack   -21,  -20 ) &    -20.4^{ -0.4}_{+  0.4} &     332861 &     1.63 \pm  0.06 &     5.35 \pm  0.48 &  4.08 &     1.66 \pm  0.05 &     5.23 \pm  0.23 &   2.6 &  0.99 \pm  0.01 \\
\lbrack   -20,  -19 ) &    -19.8^{ -0.3}_{+  0.3} &      78282 &     1.72 \pm  0.09 &     5.08 \pm  0.47 &  1.69 &     1.72 \pm  0.09 &     4.88 \pm  0.34 &   1.2 &  0.95 \pm  0.11 \\
\hline
\multicolumn{10}{c}{\mbox{Blue} \quad 0.1<\zphot<0.2}\\
\hline
\lbrack   -22,  -21 ) &    -21.1^{ -0.4}_{+  0.4} &      12753 &     1.85 \pm  0.13 &     5.70 \pm  0.83 &  0.86 &     1.85 \pm  0.16 &     5.67 \pm  0.64 &   0.6 &  1.22 \pm  0.17 \\
\lbrack   -21,  -20 ) &    -20.3^{ -0.5}_{+  0.5} &     210837 &     1.67 \pm  0.07 &     4.43 \pm  0.32 &  3.54 &     1.70 \pm  0.06 &     4.44 \pm  0.25 &   2.6 &  0.98 \pm  0.35 \\
\lbrack   -20,  -19 ) &    -19.4^{ -0.5}_{+  0.5} &     571748 &     1.57 \pm  0.08 &     4.75 \pm  0.73 & 11.72 &     1.62 \pm  0.09 &     4.45 \pm  0.42 &   6.9 &  1.04 \pm  0.14 \\
\lbrack   -19,  -17 ) &    -18.6^{ -0.4}_{+  0.6} &     198394 &     1.53 \pm  0.06 &     4.50 \pm  0.49 &  2.26 &     1.56 \pm  0.06 &     4.31 \pm  0.23 &   1.2 &  1.00 \pm  0.10 \\
\hline
\multicolumn{10}{c}{\mbox{Blue} \quad 0.0<\zphot<0.1}\\
\hline
\lbrack   -21,  -20 ) &    -20.3^{ -0.7}_{+  0.9} &      39246 &     1.61 \pm  0.14 &     4.84 \pm  0.82 &  6.52 &     1.65 \pm  0.13 &     4.66 \pm  0.31 &   3.2 &  0.97 \pm  0.10 \\
\lbrack   -20,  -19 ) &    -19.3^{ -0.7}_{+  0.9} &     109606 &     1.53 \pm  0.06 &     4.63 \pm  0.45 &  2.42 &     1.57 \pm  0.07 &     4.45 \pm  0.40 &   2.4 &  0.94 \pm  0.21 \\
\lbrack   -19,  -17 ) &    -18.1^{ -0.8}_{+  1.0} &     264699 &     1.54 \pm  0.08 &     4.16 \pm  0.63 &  7.29 &     1.58 \pm  0.11 &     3.85 \pm  0.30 &   4.4 &  0.86 \pm  0.22 \\
\lbrack   -17,  -14 ) &    -16.6^{ -0.9}_{+  1.3} &      14305 &     2.02 \pm  0.23 &     4.17 \pm  1.41 &  5.05 &     1.99 \pm  0.28 &     4.34 \pm  1.00 &   2.1 &  0.82 \pm  0.33 \\
\hline
\end{array}
\end{math}
\end{table*}

We repeat the clustering analysis splitting the samples into red and blue colour using equation~\ref{eq:colour_cut}. For each new sample we re-estimate the underlying redshift distribution used in the inversion of Limber¢s equation. The corresponding 50$^{\rm th}$, 16$^{\rm th}$ and 84$^{\rm th}$ percentiles of the underlying absolute magnitude distributions are given in Tables~\ref{table:r02} and \ref{table:r03}. We also repeat the procedure outlined in Section \ref{clust:colour_dependence} for the error estimation.  

In Fig.~\ref{fig:cor_colours} we present the angular correlation functions in each luminosity and photo-$z$ bin, for red and blue galaxies. The power law fits over approximately fixed comoving scales, their corresponding errors as well as the quality of the fits and the correlation length are estimated as in Section~\ref{clust:lum_dependence} and summarized in Tables~\ref{table:r02} and \ref{table:r03}. As noted earlier, the power law fits describe the clustering measurements quite well in a qualitative sense, although certainly not well enough in a quantitative sense, with most samples presenting a typically too large reduced $\chi^2$ (see Tables~\ref{table:r02} and \ref{table:r03}). 
  
For all absolute magnitude ranges, the red population displays a steeper correlation function slope than the blue one. Blue galaxies have a much shallower slope which gradually decreases with luminosity until a sudden increase in the slope for the faintest luminosity range probed (Table ~\ref{table:r03}). 

The correlation length of red galaxies for all redshift bins presents a minimum value around $M^*$, with increasing values both faintwards and brightwards (Table~\ref{table:r02}). We note however, that this result comes with large uncertainties.
For red galaxies the correlation lengths of the brightest and faintest bin are comparable and faint red objects are more strongly clustered than red objects with intermediate luminosities. For the blue population $r_0$ behaves more regularly (like the overall population), gradually decreasing with luminosity and redshift. Blue galaxies generally have smaller uncertainties as well. Our measurement of the correlation length for the faintest luminosity bin ($r_0 = 4.17 \pm 1.41 h^{-1}\text{Mpc}$) indicates that these galaxies are similarly clustered to blue galaxies of intermediate luminosity. The robustness of this result and some caveats are discussed in Section \ref{clust:faintblue}. 

Due to the complicated way that the slope and the correlation length, as well as their respective uncertainties, change between colour selected samples, we chose to study more quantitatively the clustering of these samples using the relative bias, i.e. their clustering with respect to the $L^*$ sample. Our relative bias results for all samples, selected by photometric redshift, absolute luminosity and colour, are presented in Section \ref{clust:relbias}.  

\subsection{Clustering of faint blue galaxies}\label{clust:faintblue}

\begin{figure}
\begin{center}
\includegraphics[width=\linewidth]{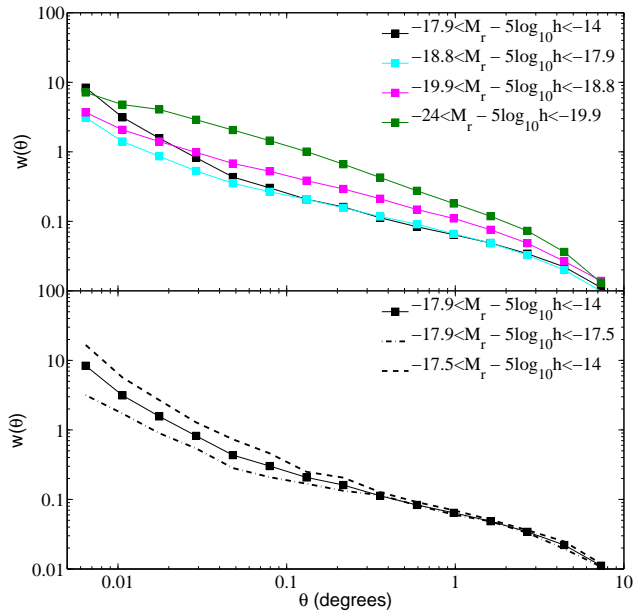}
\caption{Angular correlation functions for the low redshift galaxies in our sample split in luminosity bins. The finer luminosity binning allows one to track the scales where contamination effects (studied and quantified in Appendix \ref{clust:systematics_faint}) are significant. Error bars have been omitted for clarity.}
\label{fig:lowz}
\end{center}
\end{figure} 

One of the aims of this paper is to study the clustering of intrinsically faint galaxies for which only photometric redshifts are available in sufficient numbers to reliably calculate $w(\theta)$. The GAMA depth and the extensive SDSS sky coverage allow us to measure the auto-correlation function of the faintest optically selected galaxies, i.e. with photo-$z$ estimated absolute magnitudes in the $-17<M_r -5 \log h<-14$ range and $\zphot<0.08$. This faint sample contains a total of 14,659 galaxies, which are mostly star-forming (as evident by their colours). From the subset with spectroscopic redshifts, the 68-central percentile of the actual absolute magnitude distribution covers the range $-18<M_r -5 \log h<-12.7$. However, as shown in Appendix~\ref{clust:systematics_faint}, this sample suffers from an overall 50 per cent contamination, with most spurious objects arising from local, over-deblended spiral galaxies. 

The upper panel of Fig.~\ref{fig:lowz} shows the
correlation functions of all galaxies in our sample with $\zphot<0.08$ split into finer luminosity bins than used previously.  There exists a seemingly artificial steepening of $w(\theta)$ on scales $\theta<0.1^\circ$ for galaxies with $M_r -5 \log h>-17$.
In the bottom panel of Fig.~\ref{fig:lowz}, we further split the $-17.9<M_r -5 \log h<-14$ range into two finer luminosity bins, and again we find that for fainter samples,  source contamination affects larger angular scales. We study this contamination and quantify it as a function of scale in Appendix~\ref{clust:systematics_faint}. 

Having established the angular scales over which we trust our $w(\theta)$ measurements, we proceed to the clustering analysis. Using only the diagonal elements of the covariance matrix\footnote{Use of diagonal covariance 
elements only is appropriate for this faint sample, as it covers a rather small volume for which 
JK resampling is unable to provide an accurate description of the full covariance matrix.}, 
we note that a
power law describes the clustering signal rather well, even though
there is a hint of an increase in the clustering strength at $\sim 1 \
h^{-1}\text{Mpc}$. It is possible that this increase is due to 
blue galaxies that are satellites in small dark matter
halos. These halos should not be dense enough to stop star formation
and thus we observe only blue galaxies in this luminosity range
\citep{Celine2008}. A recent detailed study of the star formation
history of H$\alpha$ -selected faint blue galaxies in GAMA can be found in \cite{Brough2011}.  

In conclusion, the angular clustering for the faintest sample has a spurious amplitude at small angular scales, unless one takes into account the sample contamination. We do this  in Appendix~\ref{clust:systematics_faint} where we visually inspect $\sim10$ per cent of the objects in this sample and find that a significant fraction of them are spurious, mainly due to poorly deblended sources. We quantify the effect of this contamination in Appendix~\ref{clust:systematics_faint} for all luminosity bins. This investigation reveals that the angular clustering results on scales $\lesssim 0.1$ degrees are not trustworthy enough to be considered reliable. We note that the power law fits are performed on larger scales, which we show are unaffected by this contamination. However, much more detailed investigation of the data is required to robustly confirm the observed increase in the slope of the correlation function. Finally, we note that we have repeated the analysis presented in this Section for objects selected from the most recent SDSS release, DR8 \citep{Aihara11}, and we observe no differences in the results. The contamination from over-deblended spiral galaxies is still present in DR8 for the low luminosity bin. 

\subsection{Quality of fits and the HOD formalism}\label{sec:hod}
 
The power law fits presented in Table~\ref{table:r0} are not all satisfactory in a quantitative sense. The angular correlation function is only to first order well-described by a power law. The rather high reduced $\chi^2$ for some samples are either due to underestimated errors or due to the power law model being inadequate in describing the angular correlation function over a large range of scales. From the test of Section~\ref{clust:pix}, we conclude that the JK method gives consistent errors irrespective of the way we define the jackknife regions, and therefore it is most likely that the large reduced $\chi^2$ values are more due to a limitation in the power law model rather than in the error estimates themselves.

A more sophisticated model, like the halo occupation distribution (HOD) model \citep[for a review see][]{Cooray2002}, would provide a more physically motivated description of the full correlation function shape, both as a function of colour and luminosity \citep{Zehavi2004,Zheng2005,zehavi05,Zehavi2010}. 
%In this formalism galaxies are formed in virialized dark matter halos and thus the mass and the angular momentum of the parent halo determine the properties of the galaxies residing in the halo. Red galaxies which have their star formation suppressed tend to reside in dense environments, whereas blue galaxies with ongoing star formation live in less dense environments. In a similar manner, luminous galaxies reside in more massive halos and thus are more strongly clustered, since massive halos are more strongly clustered themselves. 
The HOD framework, as shown by \cite{zehavi05}, explains the increase
of clustering in the faint red population. Bright red galaxies are
central galaxies in massive halos, whereas faint red galaxies are
satellite galaxies in massive halos. 
%The issue of actual fraction of faint red galaxies that are satellites in their halos has not been conclusively settled \citep{Zehavi2010,Ross2011}. 
Our measurements suggest that both bright and faint red galaxies are more strongly clustered than red galaxies with intermediate luminosity. 
We also observe a bump in the angular correlation function of red galaxies at
separations $\sim 1 \ h^{-1}\text{Mpc}$ which signals the transition
(change in slope) between the one-halo and two-halo term in the
correlation function. On the contrary, such a change in slope is not evident for the blue population, hence they have a smaller $\chi^2_{\nu}$. This is also in agreement with HOD predictions, which predict a simple power law for blue galaxies with luminosities $M_r -5 \log h<-21$ \citep{zehavi05}. A complete HOD modelling of these angular clustering results with photometric redshifts is beyond the scope of the present work, as this would require photo-$z$ dedicated HOD tools to be developed as the standard threshold samples cannot be defined.

\section{Bias measurements}\label{clust:bias}

\subsection{Relative bias and comparison with previous studies}\label{clust:relbias}

\begin{figure}
\begin{center}
\includegraphics[width=\linewidth]{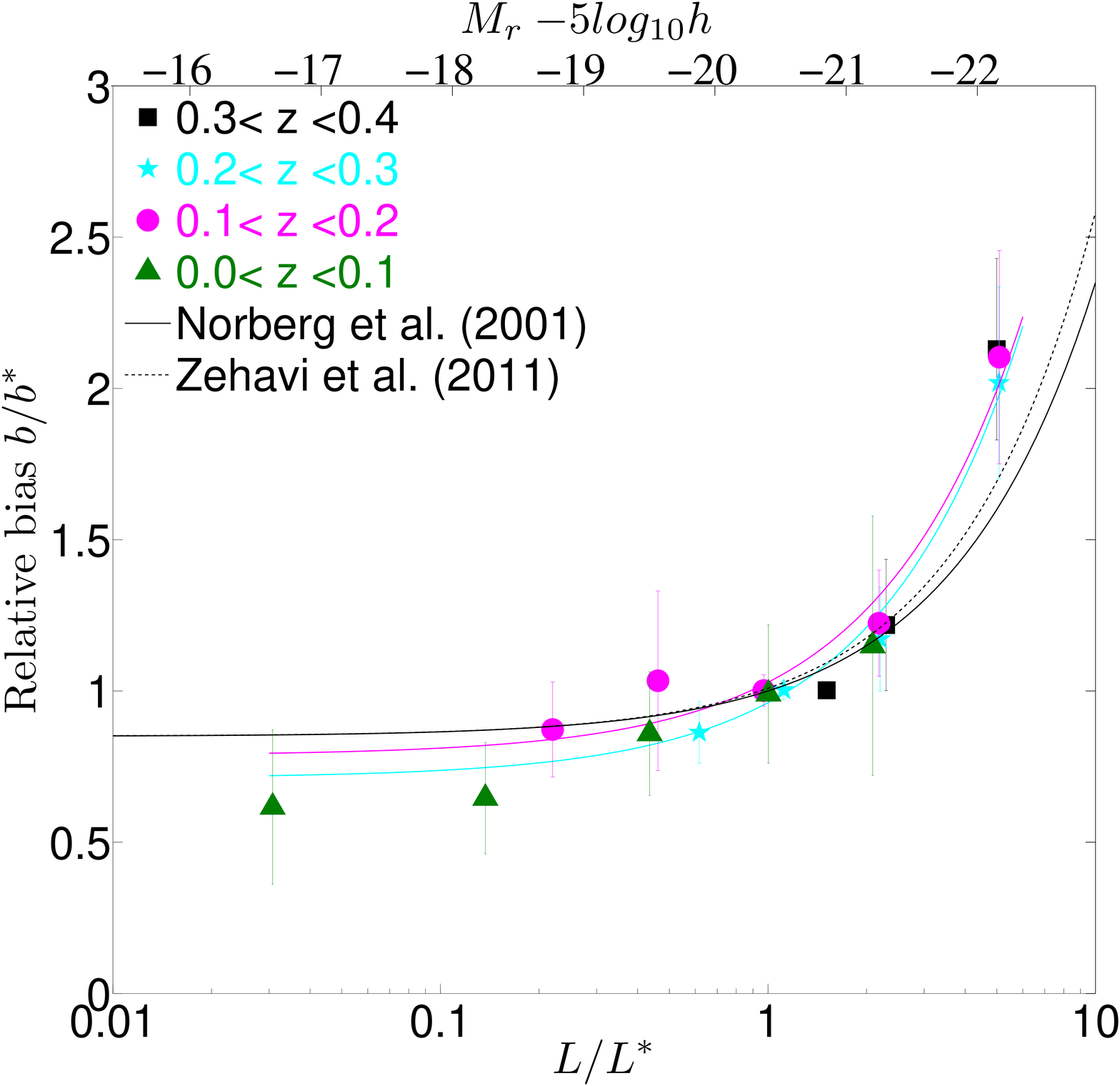}
\caption{The relative bias, defined in equation~\ref{eq:relbias}, at separations $ r = 5 \ h^{-1}$ Mpc, of all the absolute magnitude selected samples used in this study. Data points show the mean and errors of $b/b^*$ obtained from the distribution of 80 JK measurements (Sec. \ref{clust:pix}) appropriately scaled to account for the jackknife correlations. Cyan and magenta lines show our fits over the redshift ranges $0.2<\zphot<0.3$ and $0.1<\zphot<0.2$ respectively.  The solid black line shows the fit of  \citet{Norberg01} and the dotted line the fit of \citet{Zehavi2010}.}
\label{fig:relbias}
\end{center}
\end{figure}

In this paper we parametrize the real space correlation function with
a power law, and infer $\xi(r)$ from angular clustering measurements
via a Limber inversion. To ease comparison with samples using
similar, but not identical,  selection, we follow \cite{Norberg2002} and define the relative bias of a class of galaxies $i$ with respect to our $L^*$ ($-21<M_r - 5\log h<-20$) sample as
\begin{equation}\label{eq:relbias}
\frac{b_i}{b^*}(r)= \sqrt{\frac{(r_0^i)^{\gamma _i}}{r_0^{\gamma}}r^{\gamma-\gamma_i}
}.
\end{equation}
Equation~\ref{eq:relbias} preserves any scale dependence for samples
with different slopes and we choose here to estimate the relative bias
at $r = 5 \ h^{-1}$ Mpc. The advantage of using this definition of
relative bias instead of the raw correlation length to compare with
other studies is twofold. First, the former uses the slope as well as
the correlation length, which as we know from equation \ref{equ:limber_inv} are strongly correlated. Second, if the
sample selections are just slightly different, the relative bias is a much more robust way of comparing them as it measures deviations from a series of appropriate reference samples. In this study this is particularly important, as photo-$z$ inferred properties are not straightforwardly related to the underlying ones, as shown in Section~\ref{clust:dndz}. Our results are shown in Fig.~\ref{fig:relbias}.  

Previous studies from both 2dFGRS \citep{Norberg01,Norberg2002} and
SDSS \citep{Zehavi02,zehavi05,Zehavi2010} have established that the
relative bias, $b/b^*$, as a function of relative luminosity, $L/L^*$,
is well described by an affine relation. We compare our results with
these studies in Fig.~\ref{fig:relbias}. For all luminosity
bins given in Table \ref{table:r0} we fit the equation
\begin{equation}\label{eq:biasfit}
b/b^* = a_0 +a_1 L/L^*,
\end{equation} 
where $a_0$ and $a_1$ are free parameters. 
Our best fit values for samples selected on luminosity, colour and photo-$z$, using the corresponding $L^*$ for each sample, are given in Table \ref{table:bias}. The high redshift bin only provides three data points and thus we do not include it in this exercise (black squares in Fig.~\ref{fig:relbias}). In this Table we also compare with the bias relation of 
\citet{Norberg01} who found $(a_0, a_1) = (0.85,0.15)$.
The $\Delta\chi^2$ between our best fit and that of Norberg et al. is $1.2$ to $2.3$, which makes the fits statistically compatible, as the $68\%$ confidence interval for 2 degrees of freedom corresponds to $\Delta \chi^2 = 2.31$ \citep{Press1992}. 
\citet{Zehavi2010} measured the bias relative to dark matter, and in Fig.~\ref{fig:relbias} we rescale their relation with respect to $L^*$. They also observed a steeper rise in relative bias at high luminosities. Including a power of $(L/L^*)$ in our fit, we also obtain a steeper slope whilst $\chi^2$ remains unchanged, despite the additional degree of freedom. 

\begin{figure}
\begin{center}
\includegraphics[width=\linewidth]{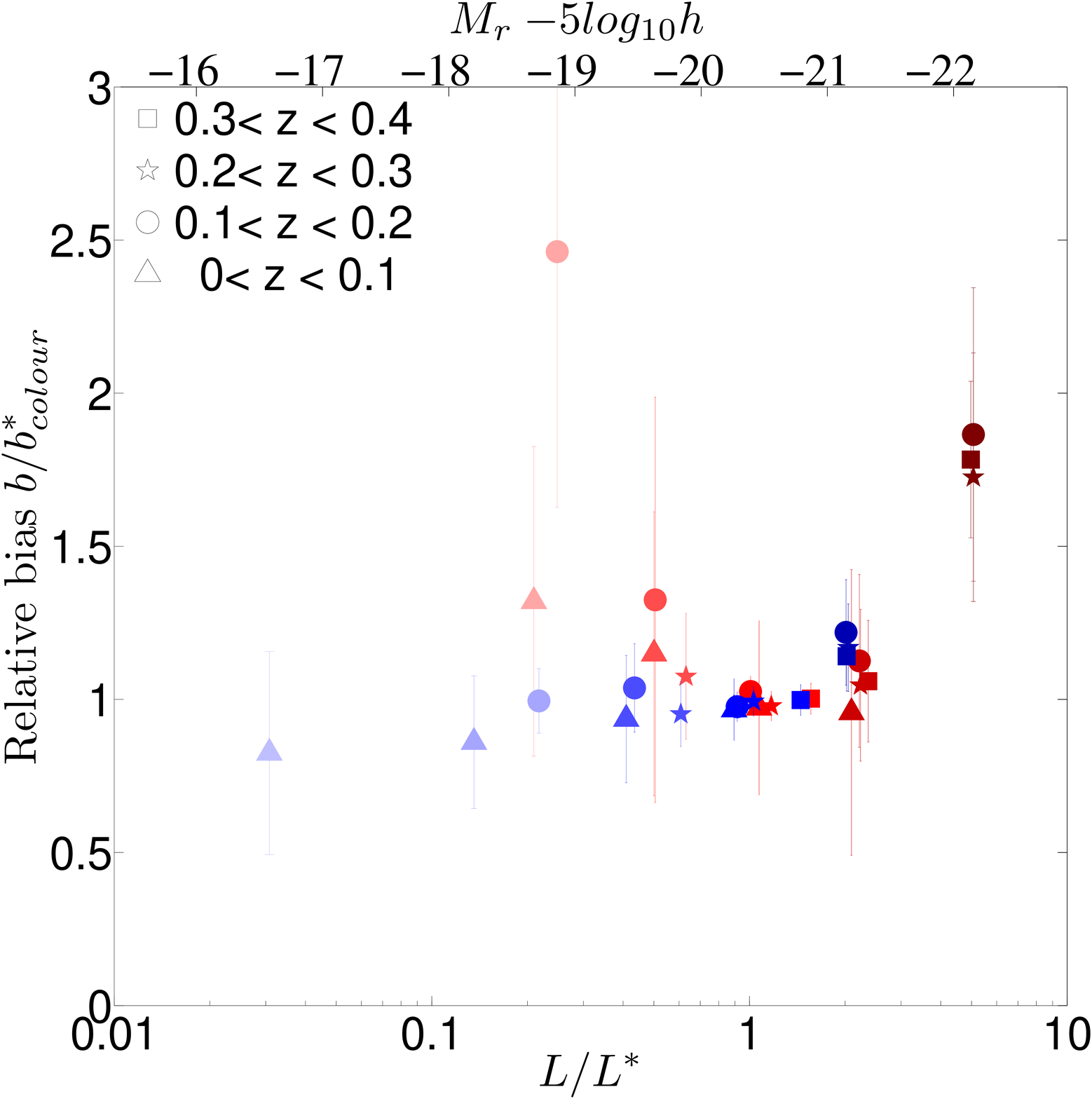}
\caption{The relative bias, defined in equation~\ref{eq:relbias}, at separations $ r = 5 \ h^{-1}$ Mpc, of all the samples used in this study split by colour (equation \ref{eq:colour_cut}). Data points show the mean and errors of $b/b^*$ obtained from the distribution of 80 jackknife measurements (Sec. \ref{clust:pix}) appropriately scaled to account for the jackknife correlations. Colour coding is as in Fig.~\ref{fig:cor_colours}. }
\label{fig:relbiascolours}
\end{center}
\end{figure}

For samples selected by colour as well as luminosity, it is more difficult to fit equation~\ref{eq:biasfit} in each redshift bin. For most photo-$z$ bins we have four or fewer data points. Moreover, using finer luminosity bins would worsen the statistical errors on $N(z)$ and $N(M_r)$ and thus make any fit more difficult to interpret. 
Fig.~\ref{fig:relbiascolours} shows that the blue population follows a similar trend to the full sample but the relative bias changes more smoothly as a function of luminosity. Table \ref{table:bias} gives the values of $a_0$ and $a_1$ for the colour selected samples. We fit the same linear relation for red galaxies as well, despite the fact that a quadratic function would seem more appropriate. $\chi^2$ values for the linear fit are also shown in Tabel \ref{table:bias} and from a purely statistical point of view, a linear relation between $b/b^*$ and $L/L^*$ is still acceptable. Fig.~\ref{fig:relbiascolours} shows that the statistical uncertainty for the two faint red samples is quite large. This is due to the small number of objects in the $-19<M_r - 5\log h<-17 $ sample and due to the poor quality of fit for the $-20<M_r - 5\log h<-19 $ sample. 

\begin{table}
\caption{Fitted values of $a_0$ and $a_1$ in the bias--luminosity relation (equation~\ref{eq:biasfit}) in three photo-$z$ ranges.  Column 1 lists the redshift bin limits, columns 2, 3 and 4 the fitted values and the quality of fit (reduced $\chi^2$) and column 5 lists  $\Delta \chi^2$ between our best fit values and the fit by \citet{Norberg01}. 
}
\label{table:bias} 
\begin{math}
\begin{array}{rrrrr}
  \hline\hline
  \multicolumn{1}{c} {\mbox{Redshift range}} & 
    \multicolumn{1}{c}  {a_0} & 
      \multicolumn{1}{c} {a_1} &
      \multicolumn{1}{c} {\chi^2_{\nu}} &
      \multicolumn{1} {c} {\Delta \chi^2} \\
      \hline
%\hline
\multicolumn{5}{c}{\mbox{All colours} }\\
\hline
0.2<\zphot<0.3 & 0.71 \pm 0.04 & 0.25 \pm 0.02 & 1.10 & 2.32\\      
0.1<\zphot<0.2 & 0.82 \pm 0.06 & 0.24 \pm 0.03 & 0.14 & 1.79\\
0.0<\zphot<0.1 & 0.65 \pm 0.05 & 0.27 \pm 0.06 & 0.12 & 1.18\\
\hline
\multicolumn{5}{c}{\mbox{Red} }\\
\hline 
0.2<\zphot<0.3 & 0.92 \pm 0.17 & 0.12 \pm 0.07 & 0.36 & 0.29\\      
0.1<\zphot<0.2 & 1.28 \pm 0.43 & 0.03 \pm 0.17 & 2.33 & 1.76 \\
\hline
\multicolumn{5}{c}{\mbox{Blue} }\\
\hline
0.2<\zphot<0.3 & 0.84 \pm 0.08 & 0.15 \pm 0.06 & 0.29 & 0.77\\      
0.1<\zphot<0.2 & 0.98 \pm 0.07 & 0.08 \pm 0.06 & 0.23 &  4.22\\
0.0<\zphot<0.1 & 0.86 \pm 0.02 & 0.08 \pm 0.02 & 0.07 & 0.02\\     
 \hline
\end{array}
\end{math}
\end{table}

\subsection{The evolution of absolute bias for $L^*$ galaxies}\label{clust:biasL}

In Section \ref{clust:relbias} we calculated the relative galaxy bias using the $L^*$ sample ($-21<M_r - 5\log h<-20$) as our reference sample. In this Section we calculate the absolute bias of the $L^*$ population defined as the mean ratio of the observed galaxy correlation function, parametrized with a \emph{power law}, over the non-linear dark matter theoretical correlation function 
\begin{equation}\label{eq:absolutebias}
b^*(r)= \sqrt{\frac{\xi_{GG}(r)}{\xi_{DM}(r)}} = \sqrt{\frac{(r_0^*)^{\gamma^*}}{r^{\gamma^*}\xi_{DM}(r)}},
\end{equation}
where $5 \ h^{-1}\text{Mpc}<r<20 \ h^{-1}\text{Mpc}$. The theoretical power spectrum $P(k)$, was obtained using  \verb+CAMB+ \citep{camb} and the halo correction recipe of \citet{Smith2003}. We then Fourier transform the non-linear $P(k)$ to obtain the real space $\xi_{DM}(r)$ using the \verb+FFTLog+ package provided by \cite{Hamilton2000}.  

Since we have correlation function measurements of the $L^*$ population for a range of redshifts we can answer the question of whether the evolution of the bias can be described by the passive evolution model introduced by \citet{TegmarkPeebles1998}:
\begin{equation}\label{eq:biasevol}
[b(z_1) - 1] D(z_1) = [b(z_2) - 1] D(z_2),
\end{equation}
where $D$ is the growth of structure \citep{peebles80} which we calculate accurately using the \verb+growl+ package by \citet{Hamilton2001}, which includes corrections to $D(z)$ due to the presence of the cosmological constant. The model described by equation~\ref{eq:biasevol} assumes that the galaxy density field linearly traces the dark matter density field and all clustering evolution comes from the growth of structure in the linear regime, i.e. no merging. It is believed that $L^*$ galaxies have undergone very little merging since $z\approx1$ \citep{Conselice09,Lotz11}. 

\begin{figure}
\begin{center}
\includegraphics[width=\linewidth]{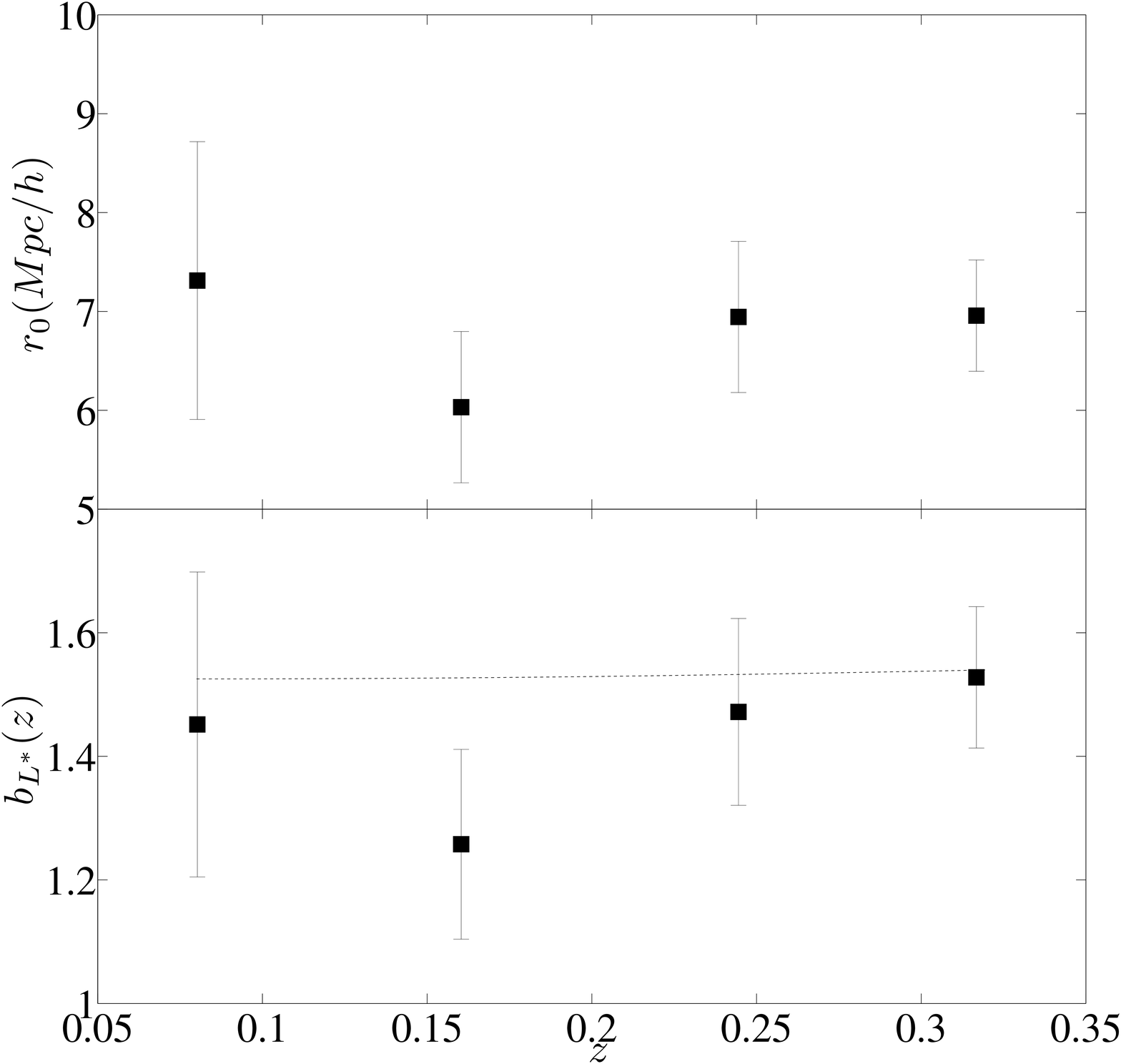}
\caption{The evolution of clustering of $L^*$ galaxies in the local universe: Upper panel shows the correlation length $r_0$; lower panel shows the bias $b_{L^*}(z)$, as a function of redshift. The dashed line in the lower panel shows the linear theory prediction from equation~\ref{eq:biasevol}. Across the redshift range $0.07 < z < 0.32$ the bias of $L^*$ galaxies agrees rather well with the linear theory model. 
}
\label{fig:absolutebias}
\end{center}
\end{figure}

In the upper panel of Fig.~\ref{fig:absolutebias} we plot the correlation length as a function of redshift. $r_0$ is observed to change very little since $z\approx0.32$. The lowest redshift point has larger errors due to the limited volume sampled. For comparisons with theory, it is more lucid to use the bias instead of the correlation length. In the lower panel of Fig.~\ref{fig:absolutebias} we plot the evolution of the absolute bias, as defined in equation~\ref{eq:absolutebias}, along with the theoretical prediction of \citet{TegmarkPeebles1998} for passive clustering evolution (dashed line). In practice, we fix the high-$z$ value of $b(z)$ and then solve equation~\ref{eq:biasevol} over the redshift range $0.07<z<0.32$. We find that the evolution of clustering of $L^*$ galaxies is consistent with the model of \citet{TegmarkPeebles1998}. 

This agreement between the clustering of $L^*$ galaxies and the passive evolution model was not observed by \citet{Ross2010} who used SDSS photo-$z$'s. The sample selection and the modeling of $w(\theta)$ and bias between this study and the one by \citet{Ross2010} are very different, as we use GAMA calibrated photo-$z$ and model the correlation function with a power law, whereas they used SDSS calibrated photo-$z$ down to $r<21$ and use halo modelling for the correlation function. Ideally one would expect that the two studies should give consistent results, but it might be that the aforementioned differences in the theoretical modelling and the sample selection influence the results significantly.   

%*****************************************************************************

\section{Discussion and conclusions}\label{clust:concl}

Despite their inherent limitations, photometric redshifts offer the opportunity to study the clustering of various galaxy populations using large numbers of objects over a wide range of angular scales with improved statistics, with the caveat that their systematic uncertainties are significantly more complex to deal with. In this section we summarize and discuss the main implications of our results.

Using GAMA spectroscopic redshifts as a training set, we have compiled a photometric redshift catalogue for the SDSS DR7 imaging catalogue with $r_{\text{petro}}<19.4$. We carried out extensive tests to check the robustness of the photo-$z$ estimates and use them for calculating r-band absolute luminosities. We split our sample of 4,289,223 galaxies into samples selected on photometric redshift, colour and luminosity and estimate their two point angular correlation functions. Redshift distributions for the Limber inversion are calculated using Monte-Carlo resampling, which we show are very reliable.  

Our clustering results are in agreement with other clustering studies such as \cite{Norberg2002} and \cite{Zehavi2010} who used spectroscopic redshifts. We extend the analysis to faint galaxies where photo-$z$s allow us to obtain representative numbers for clustering statistics. We find that the correlation length decreases almost monotonically toward fainter absolute magnitudes and that the linear relation between $b/b^*$ and $L/L^*$ holds down to luminosities $L\sim0.03L^*$. For the $L^*$ population we observe a bias evolution consistent with the passive evolution model proposed by \citet{TegmarkPeebles1998}. 

As shown by others \citep{Norberg2002,Hogg2003,zehavi05,Swanson2008,Zehavi2010} and confirmed here, the colour dependence is more intriguing because faint red galaxies exhibit a larger correlation length than red galaxies at intermediate luminosities. This trend is explained by HOD models, as shown by \cite{zehavi05}. Clustering for blue galaxies depends much more weakly on luminosity. We find that at faint magnitudes the SDSS imaging catalogue is badly contaminated by shreds of over-deblended spiral galaxies, which makes the interpretation of the clustering measurements difficult.  We determine an angular scale beyond which our results are not affected by this contamination, and test this by modelling the scale-dependance of the contamination as well as studying its luminosity dependence.    

The use of photometric redshifts is likely to dominate galaxy clustering studies in the future. A number of assumptions made in this work might need to be reviewed when we have even better imaging data and training sets. In particular, for cosmology, the non-Gaussianity of photo-$z$ and robust reconstruction of redshift distributions will become a very pressing issue. For galaxy evolution studies, it is essential to study the mapping between a photo-$z$ derived luminosity range and the true underlying one, as HOD modelling of the galaxy two point correlation function relies heavily on the luminosity range considered. In this paper, we report only qualitative agreement and leave any HOD study using these photometric redshift inferred clustering results to future work. 

%*****************************************************************************

\section*{acknowledgements}
We would like to thank the anonymous referee for comments and suggestions that significantly improved the paper. LC is financially supported by the Greek State Scholarship Foundation, trustee of the Nik. D. Chrysovergis legacy. 
CE is grateful to Gavin Dalton and Andrew Liddle for constructive comments about this work during CE's thesis defence. 
CE was partly supported by the Swiss Sunburst Fund. JL acknowledges support from the Science and Technology Facilities Council, grant numbers ST/F002858/1 and  ST/I000976/1.
SMC acknowledges the support of an Australian Research Council QEII
Fellowship and an J G Russell Award from the Australian Academy of
Science. PN acknowledges a Royal Society URF and ERC StG grant (DEGAS-259586).

GAMA is a joint European-Australasian project based around a spectroscopic campaign using the Anglo-Australian Telescope. The GAMA input catalogue is based on data taken from the Sloan Digital Sky Survey and the UKIRT Infrared Deep Sky Survey. Complementary imaging of the GAMA regions is being obtained by a number of independent survey programs including GALEX MIS, VST KIDS, VISTA VIKING, WISE, Herschel-ATLAS, GMRT and ASKAP providing UV to radio coverage. GAMA is funded by the STFC (UK), the ARC (Australia), the AAO, and the participating institutions. The GAMA website is http://www.gama-survey.org/ .

Funding for the SDSS has been provided by the Alfred P. Sloan
Foundation, the Participating Institutions, the National Science
Foundation, the US Department of Energy, the National Aeronautics
and Space Administration, the Japanese Monbukagakusho, the Max
Planck Society, and the Higher Education Funding Council for
England. The SDSS Web site is http://www.sdss.org. The SDSS is
managed by the Astrophysical Research Consortium for the
Participating Institutions. The Participating Institutions are the
American Museum of Natural History, the Astrophysical Institute
Potsdam, the University of Basel, Cambridge University, Case
Western Reserve University, the University of Chicago, Drexel
University, Fermilab, the Institute for Advanced Study, the Japan
Participation Group, Johns Hopkins University, the Joint Institute
for Nuclear Astrophysics, the Kavli Institute for Particle
Astrophysics and Cosmology, the Korean Scientist Group, the
Chinese Academy of Sciences, Los Alamos National Laboratory, the
Max Planck Institute for Astronomy, the Max Planck Institute for
Astrophysics, New Mexico State University, Ohio State University,
the University of Pittsburgh, the University of Portsmouth,
Princeton University, the US Naval Observatory, and the University
of Washington.

%******************************************************************************

% Bibliography and bibfile
\def\aj{AJ}
\def\araa{ARA\&A}
\def\apj{ApJ}
\def\apjl{ApJ}
\def\apjs{ApJS}
\def\apss{Ap\&SS}
\def\aap{A\&A}
\def\aapr{A\&A~Rev.}
\def\aaps{A\&AS}
\def\mnras{MNRAS}
\def\pasp{PASP}
\def\pasj{PASJ}
\def\qjras{QJRAS}
\def\nat{Nature}
\def\physrep{Phys. Rept.}
\def\prd{Phys. Rev. D}
\def\aplett{Astrophys.~Lett.}
\def\aas{AAS}
\let\astap=\aap
\let\apjlett=\apjl
\let\apjsupp=\apjs
\let\applopt=\ao

\bibliographystyle{mn2eNicola}

\setlength{\bibhang}{2.0em}
\setlength{\labelwidth}{0.0em}

%\bibliography{/Users/leonidas/physics/sussex/DPhil/project.bib}

\appendix

\section{SDSS SQL query}  \label{clust:query}
The SQL query used to extract our sample from the SDSS DR7 database.
\begin{verbatim}
SELECT
  objid, g.ra, g.dec, flags, petror50_r, 
  petror50Err_r, petror90_r, petror90Err_r,
  petroMag_r - extinction_r as petroMagCor_r, 
  petroMagErr_r,
  modelMag_u - extinction_u as modelMagCor_u,
  modelMag_g - extinction_g as modelMagCor_g,
  modelMag_r - extinction_r as modelMagCor_r,
  modelMag_i - extinction_i as modelMagCor_i,
  modelMag_z - extinction_z as modelMagCor_z,
  modelMagErr_u, modelMagErr_g, modelMagErr_r, 
  modelMagErr_i,
  modelMagErr_z
FROM galaxy g
JOIN Frame f on g.fieldID = f.fieldID
WHERE
 zoom = 0 and stripe between 9 and 44
 and psfmag_r - modelmag_r > 0.25 and
 petromag_r - extinction_r < 19.4
   AND ((flags_r & 0x10000000) != 0)
   AND ((flags_r & 0x8100000c00a0) = 0)
   PSF_FLUX_INTERP, SATURATED,
   AND (((flags_r & 0x400000000000) = 0) or 
   (psfmagerr_r <= 0.2))
   AND (((flags_r & 0x100000000000) = 0) or 
   (flags_r & 0x1000) = 0)
 \end{verbatim}
\section{Tests for systematics} \label{clust:systematics}

Clustering studies using photometric redshifts are subject to systematic errors which become more pressing as the statistical errors are significantly decreased. In this Appendix we study the most relevant sources of  systematic errors that might affect our results. A similar study, for a brighter sample of galaxies at higher redshifts $(0.4<z<0.7)$ was recently presented by \cite{Ross2011b}.   

Here we present tests that we believe are more likely to affect the results shown in this paper. We start in Appendix~\ref{clust:scaling} with a scaling test, which mostly tests the reliability of the whole sample for clustering studies. In Appendix~\ref{clust:crosscorr} we quantify the possible systematics in the clustering signal due to spurious cross-correlations of different photometric redshift bins. In Appendix~\ref{clust:nz} we test for possible systematics in the spatial correlation function introduced by the redshift distributions used in Limber's equation. Lastly, in Appendix~\ref{clust:systematics_faint}, we examine the robustness of the correlation function of the faintest luminosity bin.

\subsection{Scaling test} \label{clust:scaling}
\begin{figure}
\begin{center}
\includegraphics[width=\linewidth]{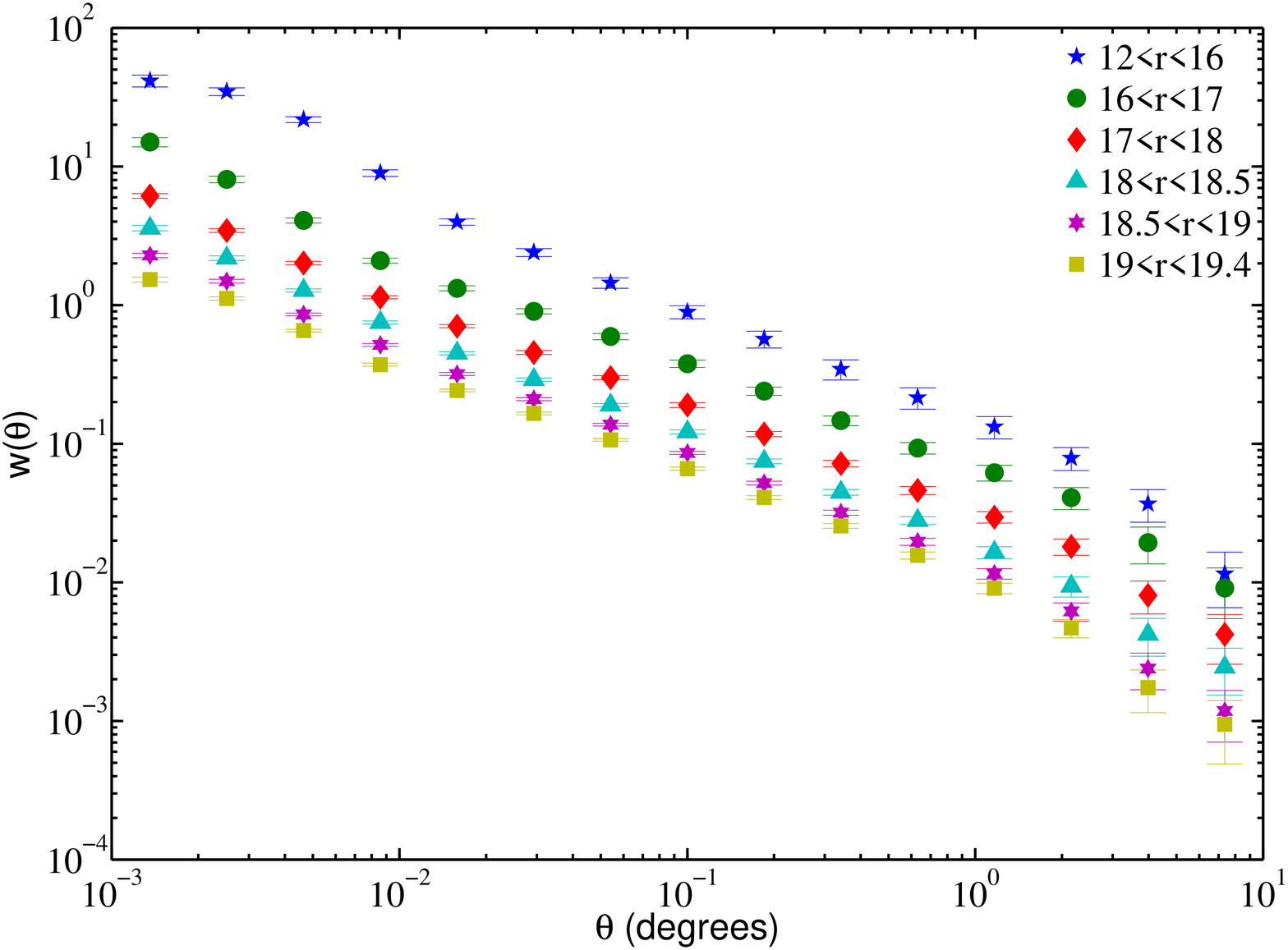}
\caption{Angular correlation functions of the r-band apparent magnitude bins defined in Table~\ref{table:rbins}.}
\label{fig:w_theta_rbins}
\end{center}
\end{figure}
\begin{table}
\caption{Clustering properties in apparent magnitude bins defined by r-band Petrosian magnitude. Column 1 lists magnitude range, column 2 the number of galaxies, columns 3 and 4 give the values of $\gamma$ and $r_0$, defined in equation~\ref{equ:powlaw_xi}. Column 5 lists the quality of the power law fits. Errors were calculated using the full covariance matrix, but we don't include the $N(z)$ uncertainty.}
\label{table:rbins}
%\vspace{3mm}
\begin{center}
\begin{math} 
\begin{array}{ccccc}
\hline \hline
\text{r-bin (mags)} & N_g & \gamma & r_0 & \chi^2_{\nu}\\ 
\hline
12.0<r<16.0 & 79543 & 1.81 \pm 0.03 & 5.01 \pm 0.48 & 1.01\\
16.0<r<17.0 & 201805 &1.72 \pm 0.02 & 5.76 \pm 0.31 & 3.1\\
17.0<r<18.0 & 671315 &1.73 \pm 0.01 & 5.62 \pm 0.20 & 3.38\\
18.0<r<18.5 & 768620 &1.74 \pm 0.01 & 5.58 \pm 0.17 & 2.28\\
18.5<r<19.0 & 1336411 & 1.73 \pm 0.01 & 5.50\pm 0.12 & 2.55\\
19.0<r<19.4 & 1720930 & 1.71 \pm 0.01 & 5.20\pm 0.12 & 3.48\\
\hline
\end{array}
\end{math}
\end{center}
\end{table}
With a photometric sample of this size it is prudent to perform a scaling test in order to uncover any dependence of clustering on apparent magnitude. In order to do this we split our sample in apparent magnitude bins and then calculate the angular correlation function. The apparent magnitude ranges are given in Table~\ref{table:rbins}. The angular correlation functions are shown in Fig.~\ref{fig:w_theta_rbins}. For all apparent magnitude bins the slope is approximately equal, but the amplitude varies as expected, shifting from high to low values as we go fainter. We then use equation~\ref{equ:limber_inv} to calculate the correlation length for each magnitude range. We fit over scales of $0.01<\theta<2$ degrees ($0.02<\theta<1.2$ degrees for the $12<r<16$ sample). The correlation length for each magnitude bin is found to be equal within the error bars and in agreement with the earlier study of \cite{Budavari2003}. Thus, for all well populated apparent magnitude bins we recover the fiducial power law \citep{peebles80}
\begin{equation}
\xi(r)\simeq\left(\frac{r}{5 \ h^{-1}\text{Mpc}}\right)^{-1.7}.
\end{equation}

\subsection{Cross correlation of photometric redshift cells}\label{clust:crosscorr}
\begin{figure*}
\begin{center}
\subfigure{
\includegraphics[width=8.5cm]{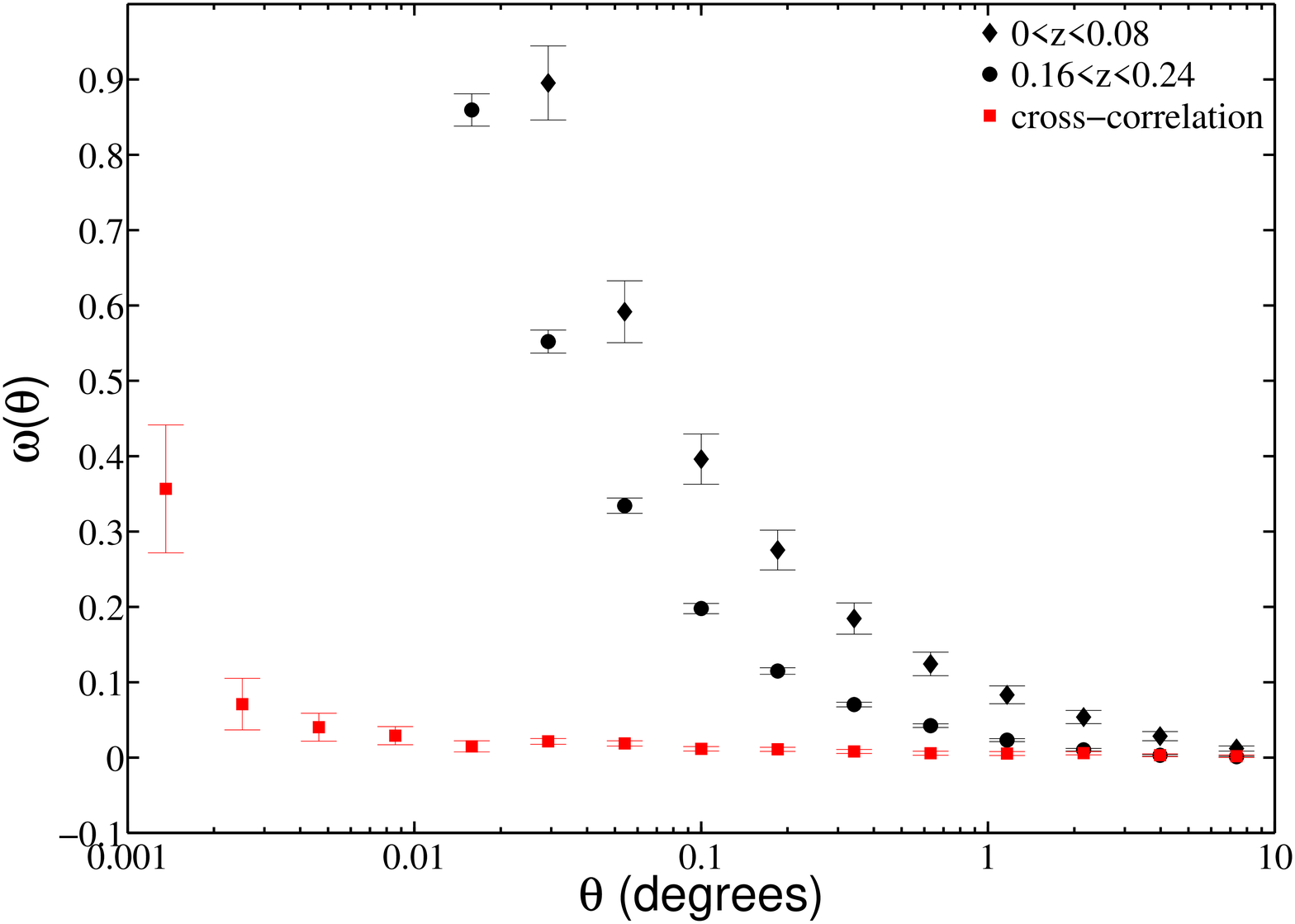}}
\subfigure{
\includegraphics[width=8.5cm]{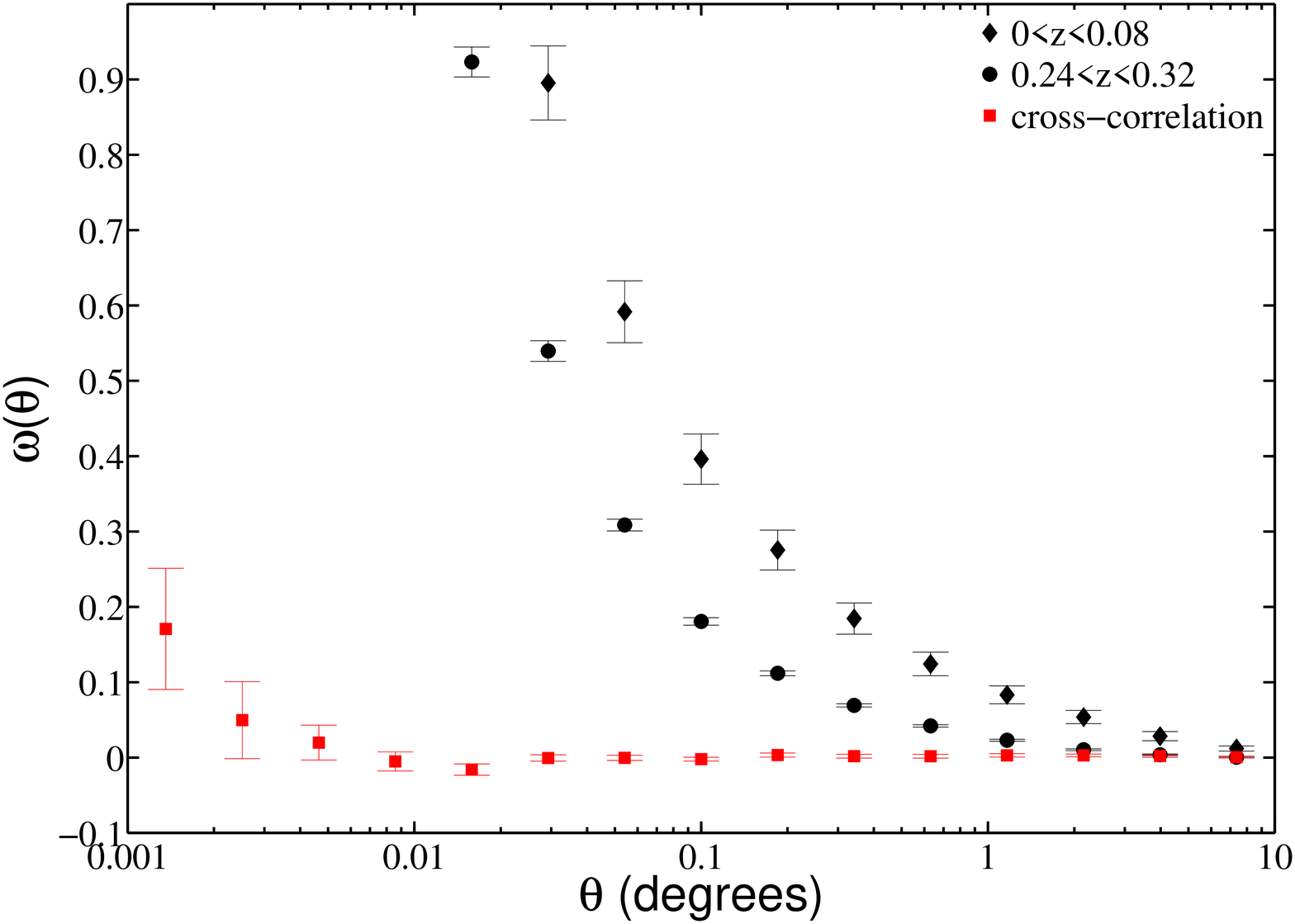}}
\\
\subfigure{
\includegraphics[width=8.5cm]{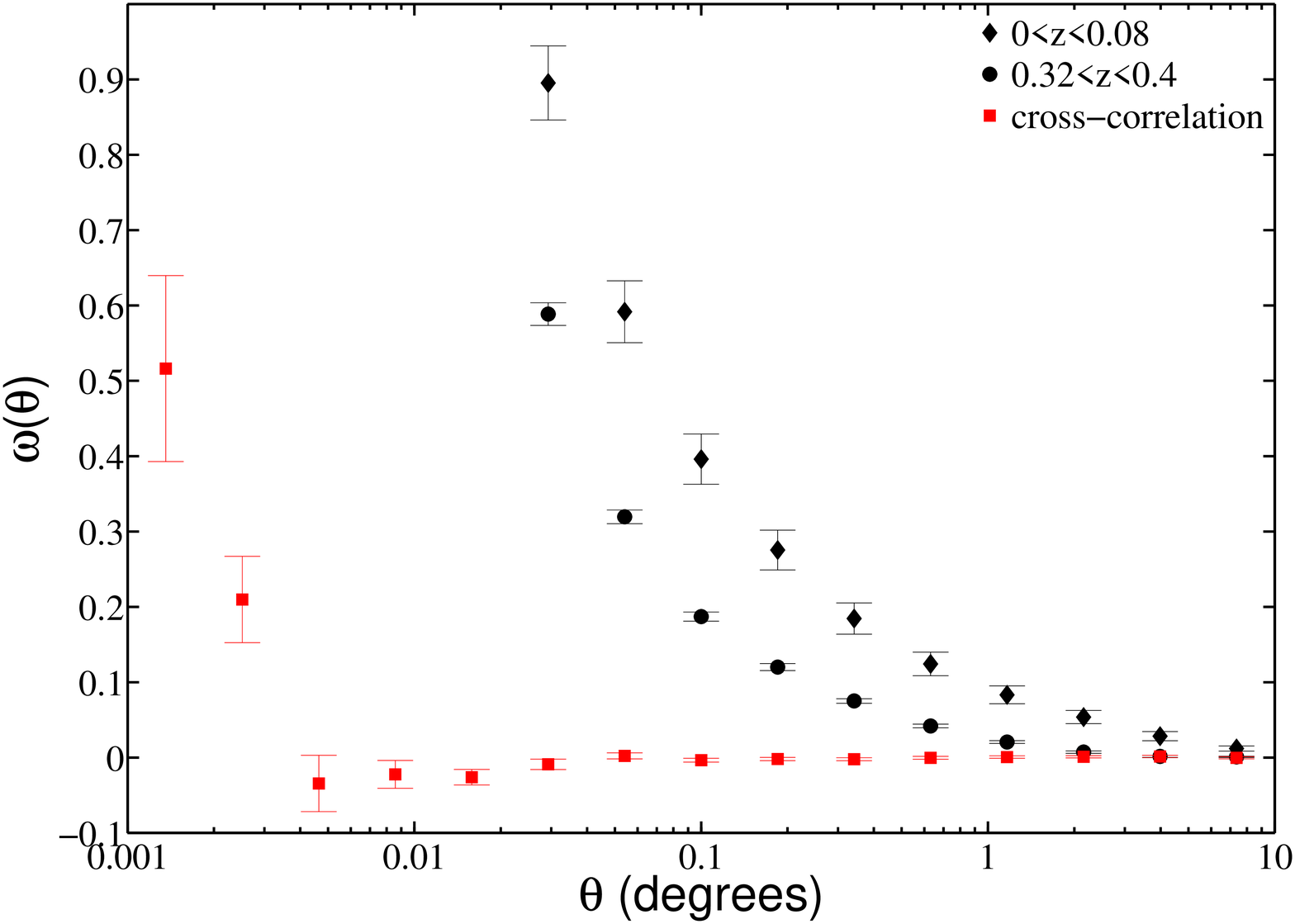}}
\subfigure{
\includegraphics[width=8.5cm]{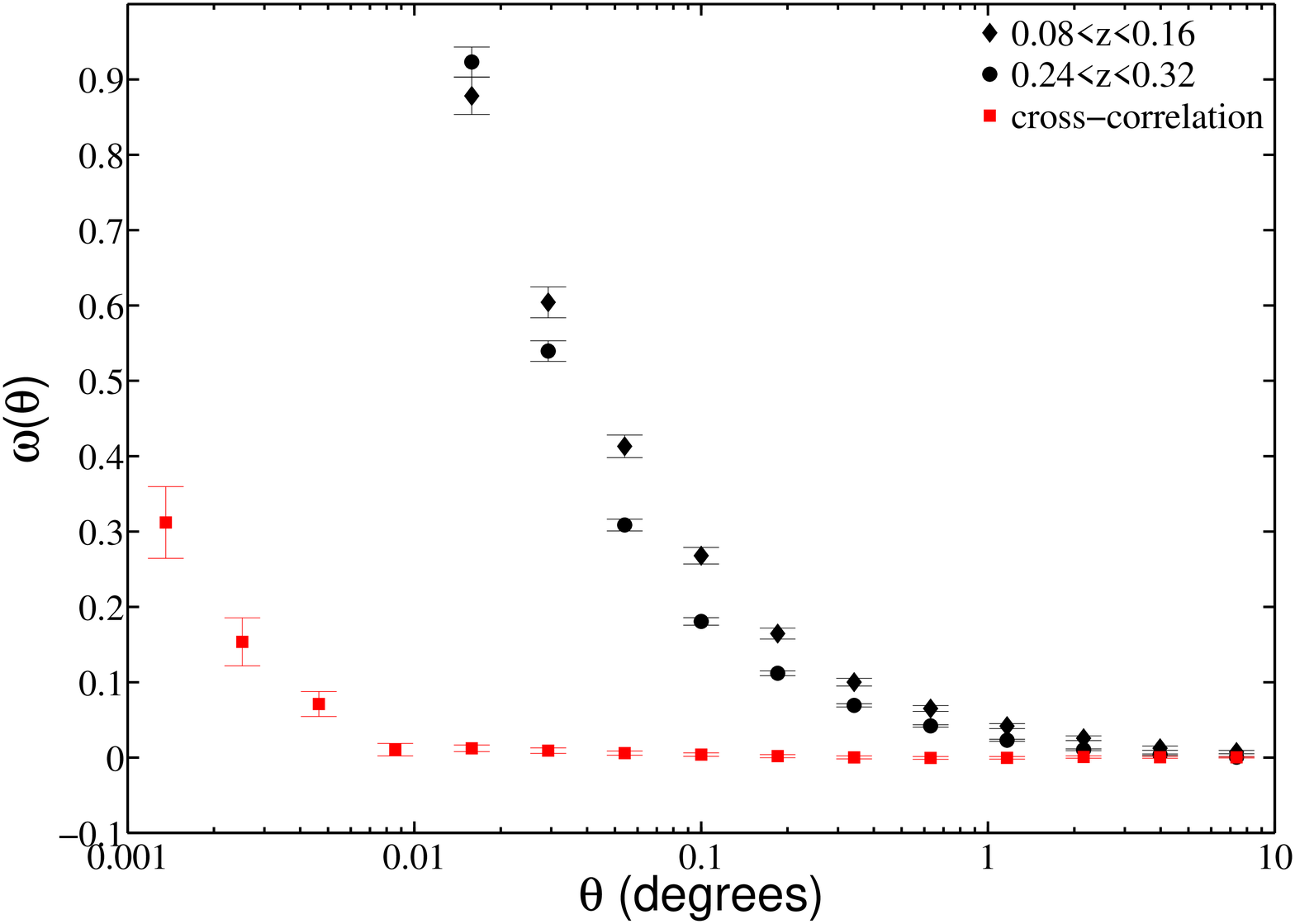}}
\\
\subfigure{
\includegraphics[width=8.5cm]{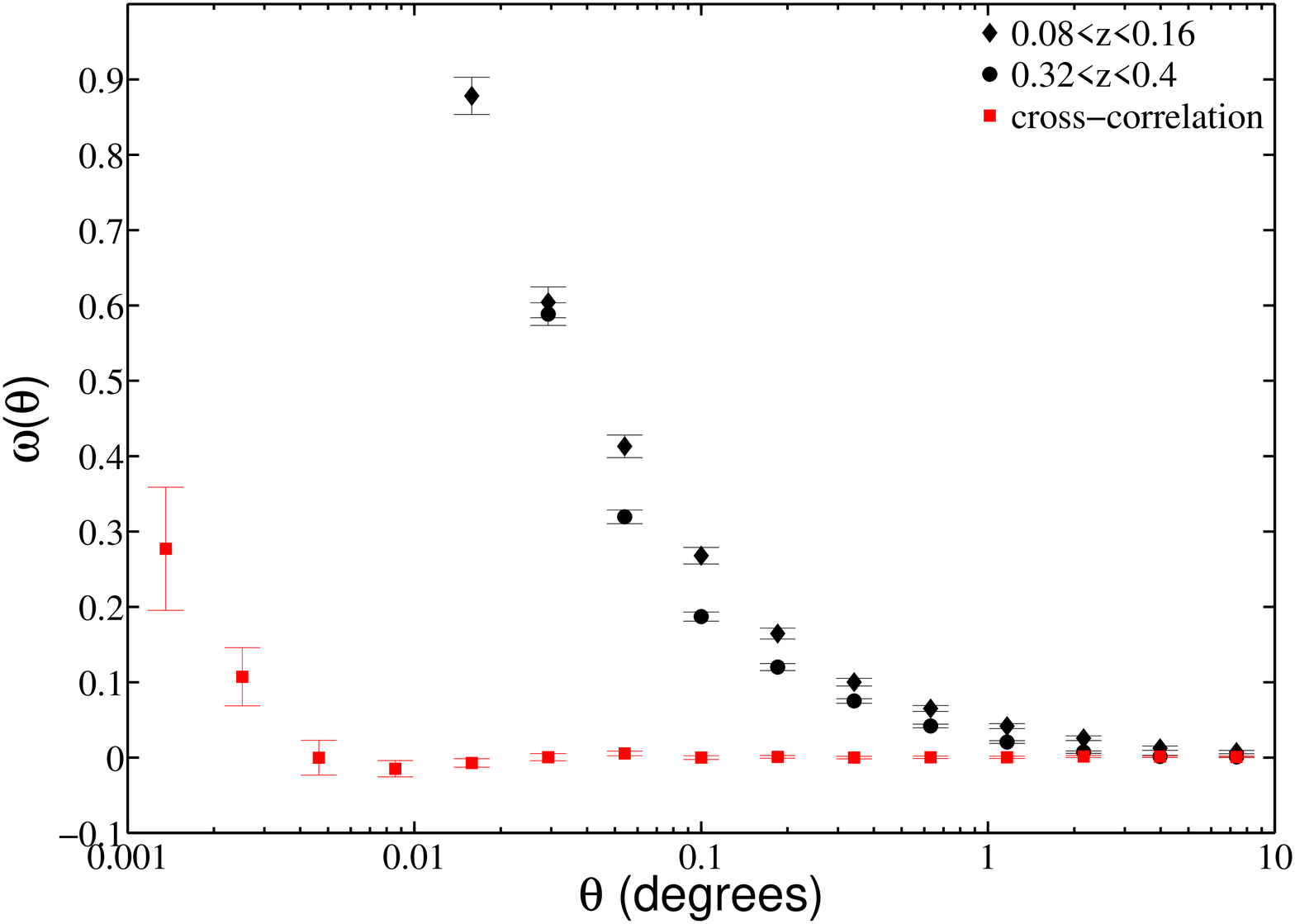}}
\subfigure{
\includegraphics[width=8.5cm]{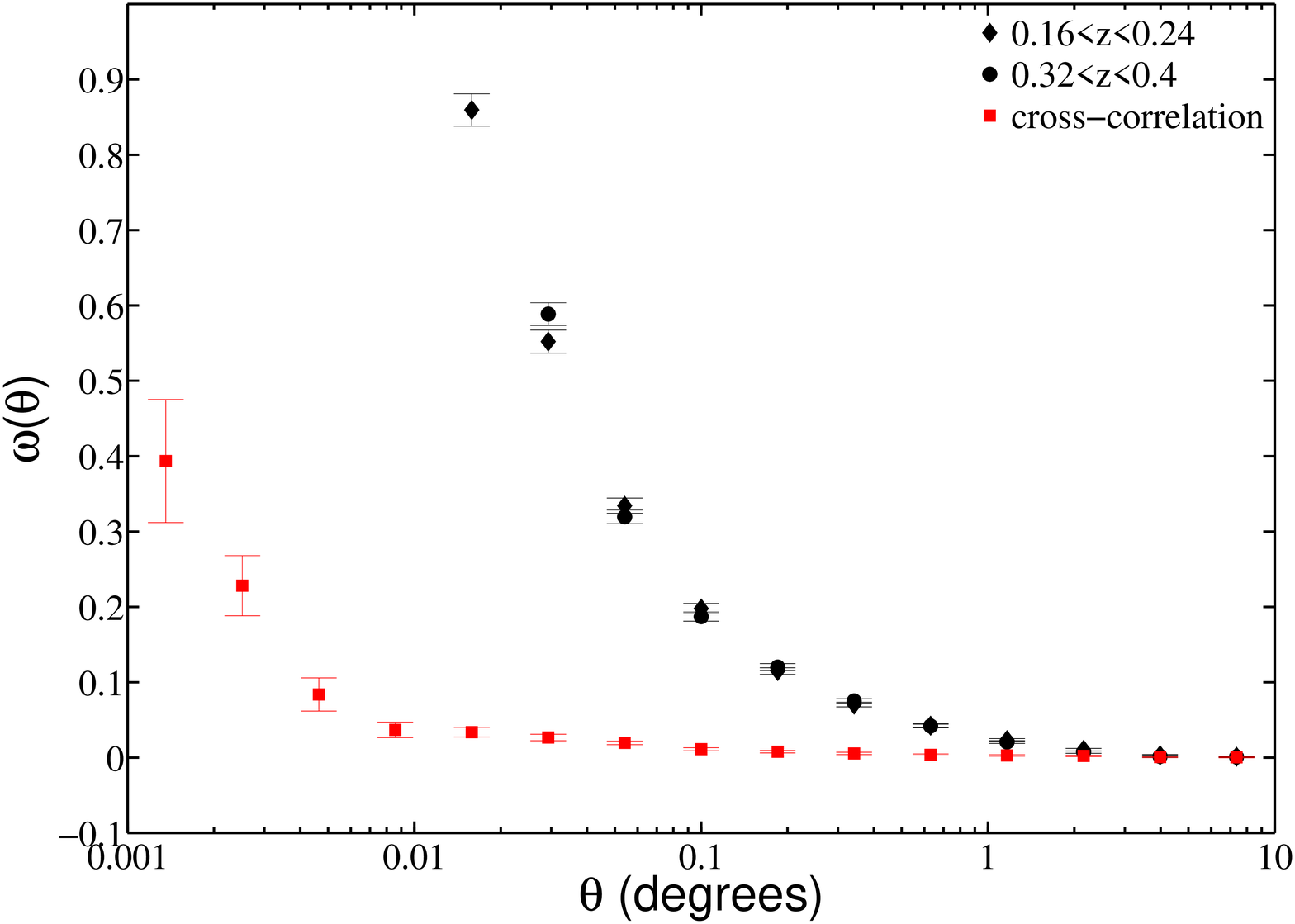}}
\\
\caption{Auto-correlation (diamonds and circles) and cross-correlation
  (squares) functions for photo-$z$ bins. The cross-correlation signals have negligible magnitude compared with the auto-correlations, and for angular separations $\geq$ 0.01 degrees are consistent with zero. The errors are calculated using JK resampling as explained in Section~\ref{clust:pix}.}
\label{fig:crosscorr}
\end{center}
\end{figure*}

A crucial consistency check, necessary for the validation of our results, is the study of the induced cross correlations between redshift shells defined by photo-$z$s from our sample. Since we have established that $\sigma_z \approx 0.04$ we start from $z_{\text{photo}}=0$ and use five continuous slices with $\Delta z = 0.08$, in order to allow all galaxies with photo-$z$ error of $\lesssim 2\sigma$ to be included in the correct redshift  bin. We then cross-correlate slices which are more than one $\Delta z$ apart. 

If a Gaussian with $\sigma=0.04$ provides good approximation of the error $\sigma_z$, then we can estimate what fraction of galaxies should lie outside the width of each photo-$z$ slice. A galaxy which is outside its redshift slice with width $\Delta z=0.08$ will have an error greater than $2\sigma$. For a Gaussian distribution $\sim5$ per cent of all galaxies should lie outside their redshift boundaries. Therefore their residual contribution to the cross correlation should be $\sim10$ per cent of their auto-correlation\footnote{Assuming that the two auto-correlations are equal and the number of galaxies in each sample is equal as well. For a detailed treatment of these effects see \citet{Benjamin2010}.}. In Fig.~\ref{fig:crosscorr} we present three auto-correlation functions and their respective cross-correlations. The cross-correlation functions from Fig.~\ref{fig:crosscorr} are not entirely consistent with zero, but on all scales the residual signal is of the expected order of magnitude. Fig.~\ref{fig:crosscorr} demonstrates that ANNz does not produce spurious correlations between physically disjoint galaxies.  

\subsection{Testing $dN/dz$}\label{clust:nz}

Here we test the accuracy of our recovered $dN/dz$ distribution by studying angular clustering in the GAMA area. Since we have precise knowledge of the spectroscopic redshift distributions in the GAMA area, we use these angular clustering measurements to test the robustness of our spatial clustering results using different methods of recovering $dN/dz$. The methods that we test against the given GAMA spectroscopic redshift distributions are (i) Monte-Carlo resampling of the photo-$z$ distributions, assuming Gaussian errors (equation~\ref{eq:noise}), which has been used for all the results in this paper, and (ii) the weighting method of \citet{Cunha2009} (also known as nearest neighbour method).

The latter method can be summed up in three distinct steps. First, one estimates the distance in apparent magnitude space to the 200th nearest neighbour of each object in the spectroscopic set, using a Euclidean metric. The exact ordinal number of the neighbouring object should not change the result significantly. For the GAMA number density, $N=200$ is the best trade-off between smoothing out the large scale structure while at the same time preserving the locality of the photometric information. Second, one calculates the number of objects in the photometric set that are within the hypervolume defined by this distance and then one calculates the weight of each object in the spectroscopic set at point $m_{i}$ according to the equation
\begin{equation}
w_{i} = \frac{1}{N_{\text{phot,tot}}} \frac{N(m_{i})_{\text{phot}}}{N(m_{i})_{\text{spec}}},
\end{equation}
where $N(m_{i})_{\text{spec}}=200$. In the third step, the already known spectroscopic distribution is weighted to match the distribution of the photometric sample. The weighting is done by summing the weights $w_{i}$ of each object in the spectroscopic sample for all redshift ranges:
\begin{equation}
N(z)_{\text{wei}} = \sum^{N_{\text{spec,tot}}}_{i=1}w_iN(z_1<z_i<z_2)_{\text{spec}}.
\end{equation}
\cite{Cunha2009} show that this method is superior in recovering the true $dN/dz$ to other methods using photo-$z$s, but they do not include the Monte-Carlo resampling in their comparisons.

\begin{figure}
\begin{center}
\includegraphics[width=1.\linewidth]{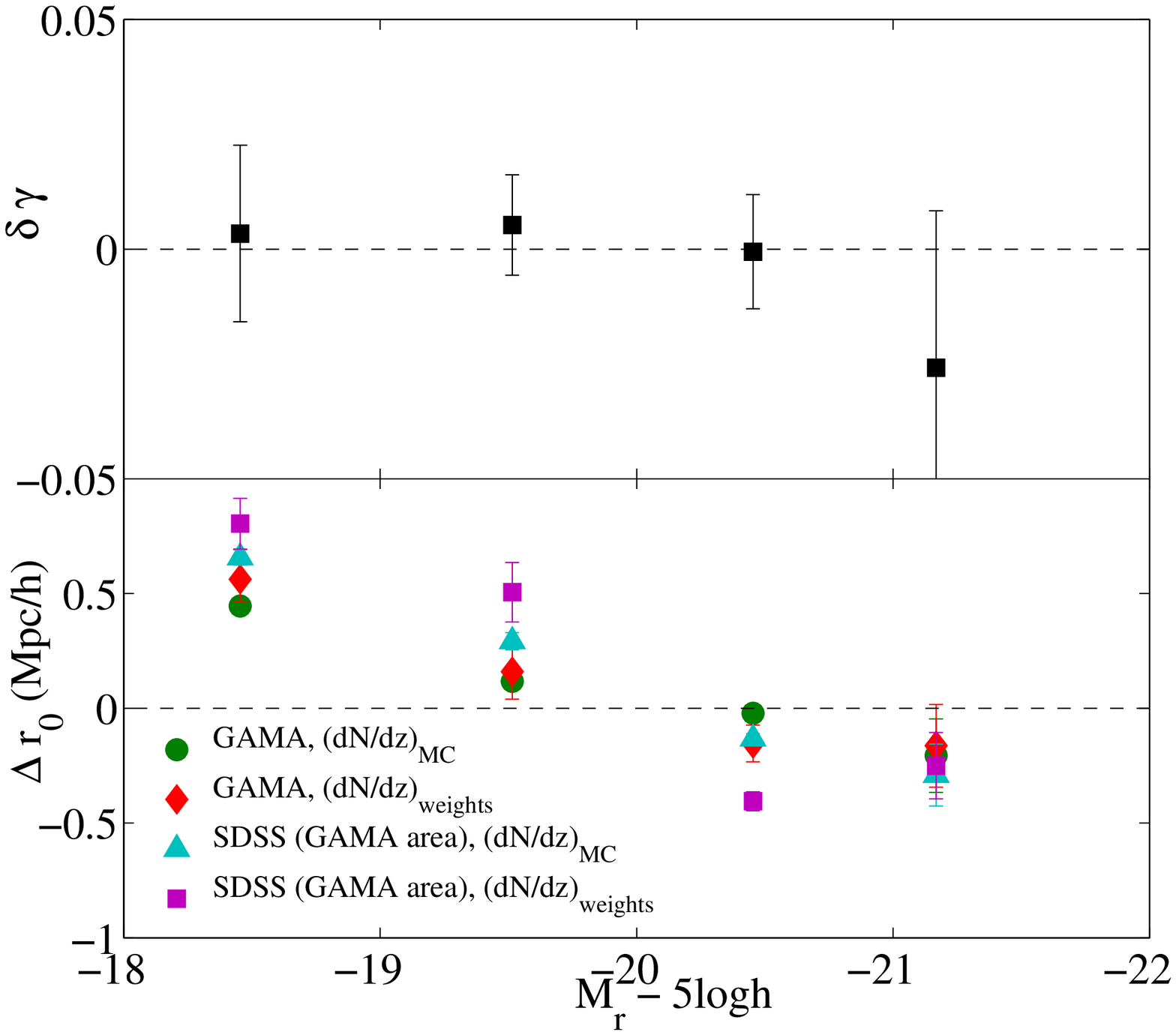}
\caption{Upper panel: Slope residual of the correlation function
  measurements in the GAMA area, using the measurement of the GAMA
  sample with spectroscopic redshifts as a reference $(\Delta \gamma =
  \gamma(\text{SDSS}) - \gamma(\text{GAMA}))$. Lower panel:
  Comparison of the effect of the various redshift distributions (as
  shown in Fig.~\ref{fig:dndz_all}) on $r_0$ measurements again using
  the GAMA sample as a reference $(\Delta r_0 = r_0(\text{i}) - r_0(\text{GAMA}))$. Following the discussion in Sec.~\ref{clust:lum_dependence}, the error bars show the combined effect of the power law fit uncertainties (assumed to be Poisson), which are independent of the underlying $dN/dz$, and the scatter in $r_0$ due to 100 Monte-Carlo resamplings of each $dN/dz$ (only $(dN/dz)_{\text{spec}}$ is known precisely). 
}
\label{fig:dndz_r0}
\end{center}
\end{figure}

The comparison of the different methods is depicted in Fig.~\ref{fig:dndz_r0}, where all the clustering measurements are confined to the GAMA area. The errors for the angular clustering measurements are assumed to be Poisson, which is just a lower bound, and the errors on the redshift distributions are obtained from the scatter of Monte-Carlo simulations. 
This test is performed for the same luminosity bins as in Section~\ref{clust:bins}, apart from the brightest and faintest bins which have a very small number of galaxies and hence large statistical errors on $w(\theta$). 

The (a priori required) agreement between the $r_0$ measurements from the different methods of recovering $dN/dz$ is not perfect. The $r_0$ measurements are not significantly affected by the differences between the redshift distributions of Fig.~\ref{fig:dndz_all}. 
In conclusion, Fig.~\ref{fig:dndz_r0}, for the three intermediate and well populated luminosity bins, implies that the reconstruction of the underlying redshift distribution is not introducing any systematic errors in the $r_0$ measurements.   

This comparison does have its limitations. Samples with small numbers of
objects are sensitive to number variations due to the different
selections of the two surveys (mainly the more conservative
star-galaxy separation that we use in this paper). Moreover, it is
very difficult to get realistic error bars for samples with a small
number of galaxies and for which the survey's angular extent is
comparable with the angular scales used for the $w(\theta)$
measurements. The difficulty in getting the exact angular clustering signal is shown in the upper panel of Fig.~\ref{fig:dndz_r0} which shows the residuals of the measured slopes for the GAMA and SDSS samples. In spite of these, Monte-Carlo resampling seems to recover the true $r_0$ slightly better than the weighting method.

\subsection{Correlation function for faint galaxies}\label{clust:systematics_faint}

\begin{figure}
\begin{center}
\label{fig:faint_blue_clean}
\includegraphics[width=\linewidth]{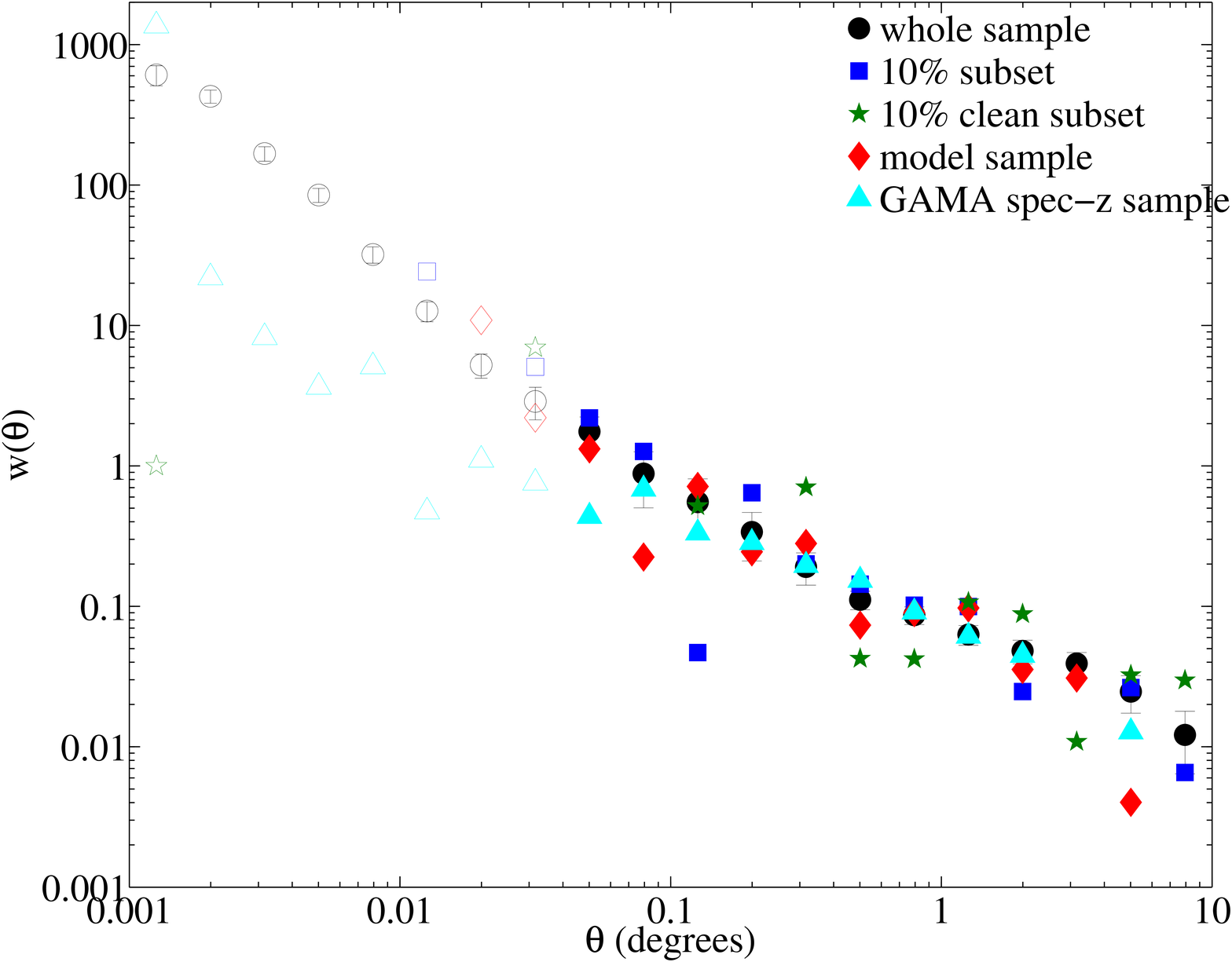}
\caption{Two point correlation function of the faintest luminosity
  bin ($-17<M_r - 5\log h<-14$). Black circles show the total correlation function, blue squares
  show the correlation function of the $\sim10$ per cent subset of
  objects visually inspected, green stars show the correlation
  function of the ``clean'' part of the inspected subset, red diamonds show the total correlation function corrected to account for the spurious pairs on scales $\gtrsim0.1$ degrees and finally, cyan triangles show the $w(\theta)$ measurement using only GAMA spectroscopic data. Errors bars for the total sample are calculated using the JK method. Open symbols represent angular scales at which the signal is significantly contaminated and so cannot be trusted. }
\end{center}
\end{figure}

The correlation function of the faintest sample [$-17$, $-14$) exhibits an infeasibly large clustering amplitude at small scales (Fig.~\ref{fig:faint_blue_clean}). This increase in the clustering signal is not hinted at in the $-19<M_r - 5\log h<-17$ luminosity bin, and so we here investigate whether there is some sort of contamination in the faintest sample.  

\begin{figure}
\begin{center}
\includegraphics[width=1\linewidth]{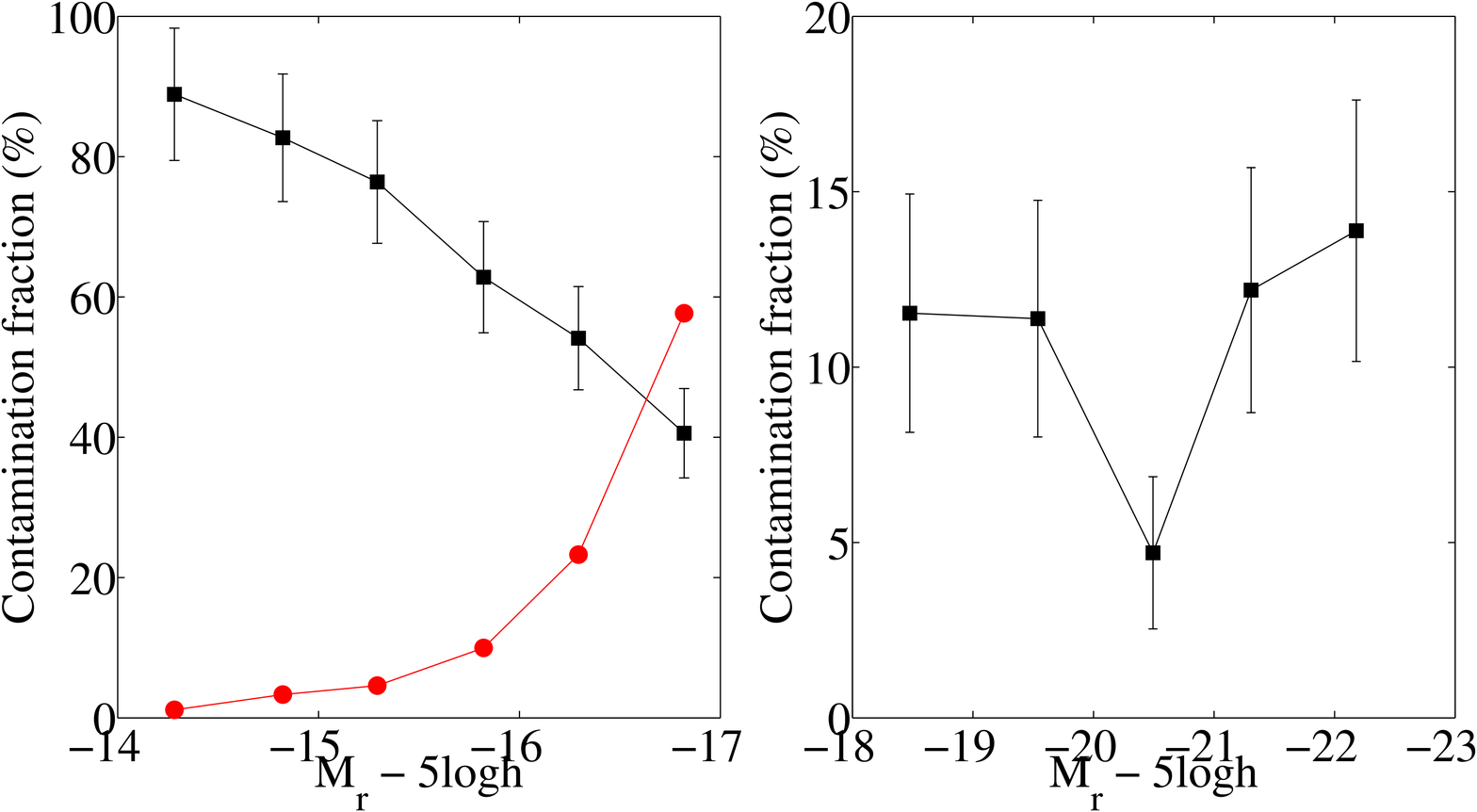}
\caption{Left panel: Black symbols show the fraction of spurious objects for the faintest luminosity bin as a function of absolute luminosity. These fractions are estimated by visually inspecting $\sim10$ per cent of the total number of objects in that bin. Red symbols show the overall distribution of objects as a function of absolute magnitude. Right panel: Fraction of spurious objects as a function of absolute luminosity, obtained by visually inspecting a small subset ($\sim100$) of all objects in each luminosity bin.
In both panels the error bars are obtained assuming Poisson statistics.}
\label{fig:faint_blue_cont2}
\end{center}
\end{figure}

We randomly select $\sim10$ per cent of the objects in the faintest
luminosity bin and we visually inspect them to see if they are genuine
galaxies. The fraction of spurious objects is shown in the left panel
of Fig.~\ref{fig:faint_blue_cont2} and we observe that it is
significant at the very faint end, where the actual number of galaxies
is low (red line in the same figure), and $\sim40$ per cent at the
bright end of that luminosity bin. From our visual inspection, most
spurious objects are local, over-deblended spiral galaxies, the remainder are merging systems or just sky noise. Evidently as we go fainter, the contamination level is increasing and this presents a serious drawback for clustering studies and a serious limitation for large surveys. 

The right panel of Fig.~\ref{fig:faint_blue_cont2} shows the fraction
of spurious objects in the other five absolute magnitude bins. We
visually inspected $\sim100$ objects from each of those bins and we
found that the contamination level is much lower, with a slight
increase toward the bright and faint ends. Our detailed study of the correlation
function of the faintest bin shows that it is not affected by
contamination on the scales of primary interest ($\theta \gtrsim 0.1^\circ$), something which we
expect to hold true for all other luminosity bins, 
which have a significantly smaller fraction of spurious objects.

The contamination in the $-17<M_r - 5\log h<-14$ luminosity bin affects the two point correlation function differently at different angular scales. We address this issue by counting the number of pairs of genuine galaxies in the visually inspected subset. The results are shown in Fig.~\ref{fig:faint_blue_clean}, where we also include the angular correlation function from the corresponding sample from GAMA.\footnote{GAMA objects have been visually inspected and are therefore more reliable than objects in the SDSS imaging catalogue. On the other hand, GAMA has a smaller area, which increases the statistical errors. For this sample, considering Poisson errors only, the statistical errors on $w(\theta)$ would be at least three times larger than the ones obtained from the SDSS sample.} Due to the fact that the subset has a weakened signal at very small scales we can only draw conclusions for angular scales $>0.1$ degrees. From Fig.~\ref{fig:faint_blue_clean} we see that at these scales the contamination does not significantly affect the correlation function and its fit parameters $\gamma$ and $r_0$.
For this reason, we present our results limited to angular scales $\theta \gtrsim 0.1^\circ$.

We also repeated our analysis after masking out areas of sky covered by RC3 galaxies \citep{RC3,RC3b} to test whether we could decrease the contamination level. We did not observe any qualitative differences in the power law parameters estimated, and more importantly, the amplitude of $w(\theta)$ at small scales did not reduce, indicating that the RC3 catalogue does not capture all over-deblended galaxies in the SDSS galaxy catalogue. 

Finally, it is important to note (and caution) that the source
contamination due to over-deblending only became apparent when interpreting
the bottom right panels of Figs.~\ref{fig:cor_absmagr}
and~\ref{fig:cor_colours}). Had we completely trusted the results of
the scaling test (Appendix~\ref{clust:scaling}) or used only the data
point near $L^*$ in Fig.~\ref{fig:faint_blue_cont2} (since that
population dominates), we would have significantly underestimated the number of spurious objects.   

\end{document}